\documentclass[journal,comsoc]{IEEEtran}
\usepackage[T1]{fontenc}% optional T1 font encoding
\usepackage{color}
\usepackage{amsmath}
\usepackage{url}
\usepackage{array}
\usepackage{booktabs}
\usepackage{fancyhdr}
\usepackage{latexsym}
\usepackage{diagbox}
\usepackage{epsfig}
\usepackage{epstopdf}
\usepackage{amssymb}
\usepackage{amsfonts}
\usepackage{amsxtra}
\usepackage{xspace}
\usepackage{makeidx}
\usepackage{graphics}
\usepackage{theorem}
\usepackage{subfigure}
\usepackage{multirow}
\usepackage{verbatim}
\usepackage{algorithmic,algorithm}
\DeclareMathOperator*{\minimize}{minimize}
\DeclareMathOperator*{\maximize}{maximize}
\usepackage[noadjust]{cite}
\newtheorem{lemma}{\textbf{Lemma}}

\begin{document}

\title{Beamforming for PIN Diode-Based IRS-Assisted Systems Under a Phase Shift-Dependent Power Consumption Model}

\author{Qiucen~Wu,~\IEEEmembership{Graduate~Student~Member,~IEEE,}
        Tian~Lin,\IEEEmembership{}
        Xianghao~Yu,~\IEEEmembership{Member,~IEEE,}
        ~Yu~Zhu,~\IEEEmembership{Senior~Member,~IEEE}
        and~Robert~Schober,~\IEEEmembership{Fellow,~IEEE}% <-this % stops a space
%\thanks{This work was supported by National Natural Science Foundation of China under Grant No. 61771147. \textit{(Corresponding author: Yu Zhu; Xianghao Yu.)}}
\thanks{%This work was supported by the Natural Science Foundation of Shanghai under Grant No. 23ZR1407300 and the China Scholarship Council. 
An earlier version of this paper was presented in part at the IEEE Global Communications Conference, Kuala Lumpur, Malaysia, in December 2023 [DOI:  10.1109/GLOBECOM54140.2023.10437013]. \textit{(Corresponding authors: Yu Zhu; Xianghao Yu.)}}
\thanks{Qiucen Wu, Tian Lin, and Yu Zhu are with the Key Laboratory for Information Science of Electromagnetic Waves (MoE), School of Information Science and Technology, Fudan University, Shanghai 200433, China (e-mail: qcwu21@m.fudan.edu.cn, lint17@fudan.edu.cn, zhuyu@fudan.edu.cn).

Xianghao Yu is with the Department of Electrical Engineering, City University of Hong Kong (CityU), Hong Kong (e-mail: alex.yu@cityu.edu.hk).

Robert Schober is with the Institute for Digital Communications, Friedrich-Alexander-University Erlangen-Nuremberg (FAU), 91054 Erlangen, Germany (e-mail: robert.schober@fau.de).}

}

\maketitle

\begin{abstract}
Intelligent reflecting surfaces (IRSs) have been regarded as a promising enabler for future wireless communication systems due to their capability of customizing favorable propagation environments. In the literature, IRSs have been considered power-free or assumed to have constant power consumption. However, recent experimental results have shown that for positive-intrinsic-negative (PIN) diode-based IRSs, the power consumption dynamically changes with the phase shift configuration, which implies that the beamforming quality of the IRS depends on the available power. Therefore, this \emph{phase shift-dependent} power consumption (PS-DPC) introduces a challenging power allocation problem between the base station (BS) and the IRS, aiming to balance the BS transmit power and the IRS beamforming quality during system design. To tackle this issue, in this paper, we investigate a rate maximization problem for IRS-assisted systems under a practical PS-DPC model. For the single-user case, we propose a generalized Benders decomposition-based beamforming method to maximize the achievable rate while satisfying a total system power consumption constraint. Moreover, we propose a low-complexity beamforming design, where the powers allocated to BS and IRS are optimized offline based on statistical channel state information. Furthermore, we extend the beamforming design to the multi-user case, where we solve an equivalent weighted mean square error minimization problem with two different joint power allocation and phase shift optimization methods. {\color{black}Simulation results indicate that compared to baseline schemes, our proposed methods can flexibly optimize the power allocation between BS and IRS, thus achieving better performance. The optimized power allocation strategy strongly depends on the system power budget. Specifically, when the provided system power budget is high, the PS-DPC is not the dominant factor in the system power consumption, allowing the IRS to turn on as many PIN diodes as needed to achieve high beamforming quality. When the system power budget is limited, however, more power tends to be allocated to the BS to enhance the transmit power, resulting in a lower beamforming quality at the IRS due to the reduced PS-DPC budget.} %When the system power budget is limited, however, the proposed methods tend to allocate more power to the BS to enhance transmitted power, while sacrificing the beamforming quality at the IRS by reducing the PS-DPC budget. %Notably, in such scenarios, a larger IRS receives even less PS-DPC budget compared to a smaller IRS.

\end{abstract}


\IEEEpeerreviewmaketitle

\section{Introduction}\label{sec:introduction}

Intelligent reflecting surfaces (IRSs) have recently attracted increasing attention for the development of next-generation mobile communication systems, as they provide a new communication paradigm to reconfigure the wireless propagation environment \cite{2020_TCCN_Li_RIS_review}, \cite{2021_TC_Zhang_IRS_tutorial}. In particular, IRSs are a type of reconfigurable metasurface with a large number of scattering elements. By smartly designing the phase shift of each element, IRSs can flexibly direct the reflected beams into the desired directions, thereby enhancing the coverage and throughput of wireless communication systems \cite{2020_JSAC_Renzo_Smart_radio_environments_review}. Owing to this promising capability, IRSs have been employed for different purposes in communication systems, e.g., multiple-input multiple-output (MIMO) communications \cite{2019_TWC_IRS_MIMO_1}, \cite{2021_TWC_IRS_MIMO_2}, unmanned aerial vehicle communication \cite{2021_WC_IRS_UAV_1}, \cite{2021_TWC_IRS_UAV_2}, and physical layer security \cite{2021_JSAC_IRS_security_1}, \cite{2019_WCL_IRS_security_2}.
%More importantly, scattering elements in IRSs are implemented by passive hardware components, which means they only require a small amount of power for operation \cite{2021_TC_Yu_green_com}.

Although there are several works evaluating the energy efficiency of IRS-assisted systems \cite{2019_TWC_Yuen_RIS_EE}, \cite{2022_TWC_Cui_EE_disctributeD_IRS}, the power consumption of the IRS itself is not fully understood, yet. In most of the existing literature, IRSs are assumed to not consume any power or have a \emph{constant} power consumption that is independent of the phase shift configuration of the IRS. For instance, the authors in \cite{2019_TWC_Yuen_RIS_EE} investigated the energy efficiency optimization of an IRS-assisted system, where the power consumption of the IRS was modeled as a function of the number of scattering elements and their bit resolution, and assumed to be independent of the phase shift configuration. The resulting constant IRS power consumption model was also employed in \cite{2021_TSP_gxiqigao_RE_maximization} to maximize the resource efficiency, which is a performance metric that reflects the trade-off between energy and spectral efficiency. The authors of \cite{2022_TWC_Cui_EE_disctributeD_IRS} further studied a system with distributed IRSs, where each IRS can be switched on or off to maximize energy efficiency. It was assumed that the IRSs in the on-state can serve the users by reflecting impinging signals, yet with a constant power consumption. The IRSs in the off-state, on the other hand, did not reflect signals, and were assumed to not consume any power. In addition, the authors of \cite{2022_TGC_Ntontin_charging_power} considered the charging and discharging power needed for state switching of the scattering elements, and investigated signal-to-noise ratio (SNR) maximization and energy harvesting in IRS-assisted systems. However, in the above works, the power consumption of the IRSs was assumed to be \emph{independent} of its phase shift design. Hence, the IRS elements can be arbitrarily reconfigured without affecting the system power consumption.
%In \cite{2021_TWC_Renzo_EE_channel_estimation}, the power consumption of channel estimation was further introduced to obtain a more comprehensive power consumption model for IRS-assisted systems. 
% The authors of \cite{2021_TWC_Renzo_EE_channel_estimation} expanded the model to include the power consumption of channel estimation and optimal beamforming feedback to the IRS. 

Contrary to the \emph{phase shift-independent power consumption} model in the literature, recent experimental results indicate that for the most prevailing types of IRSs, e.g., positive-intrinsic-negative (PIN) diode-based IRSs, the power consumption varies \emph{dynamically} depending on the states of the PIN diodes \cite{2021_OJCS_Lerosey_IRS_demo},  \cite{2022_Arxiv_Cui_dynamic_power_consumption}, \cite{2023_Arxiv_Jin_dynamic_power_consumption}\footnote{In this paper, we focus on the power consumption model and system optimization of PIN diode-based IRS-assisted systems. It is important to note, however, that several other types of IRSs exist, e.g., varactor-diode-based IRSs, RF switch-based IRSs, etc. \cite{2022_Arxiv_Cui_dynamic_power_consumption}, \cite{2023_Arxiv_Jin_dynamic_power_consumption}. Investigating the power consumption models and the resulting system optimization problem for these other types of IRSs presents an interesting topic for future research.}. This phenomenon is primarily due to the disparate power requirements of PIN diodes in the on- and off-states. In particular, a single on-state PIN diode consumes $12.6\ \mathrm{mW}$ \cite{2022_Arxiv_Cui_dynamic_power_consumption}, whereas it consumes no power in the off-state. Therefore, the \emph{dynamic power consumption} of an IRS is proportional to the number of PIN diodes in the on-state. Since the PIN diodes control the phase shift design of the scattering elements, this dynamic power consumption is strongly related to the phase shifts of the IRS elements. Hence, in the remainder of this paper, we refer to this dynamic power consumption as \emph{phase shift-dependent power consumption (PS-DPC)}. 

{\color{black}Given the PS-DPC, the beamforming quality of the IRS depends on the available power. To provide high-quality beamforming, the IRS may need to turn on many PIN diodes, thus resulting in a high PS-DPC. However, if the PS-DPC budget is restricted, most PIN diodes have to be turned off, which will seriously affect the degrees of freedom (DoFs) for IRS beamforming. Thus, it is necessary to revisit IRS beamforming design to achieve a good balance between the beamforming quality and the PS-DPC.}

%Hence, in the remainder of this paper, we refer to this dynamic power consumption of the IRS as \emph{phase shift-dependent power consumption (PS-DPC)}. 
%have not considered the phase shift-dependent power consumption (PS-DPC) of PIN diodes in IRSs, but with the assumption that the phase shifts of IRS elements can be arbitrarily reconfigured without affecting the IRS power consumption. In contrast, in recent experimental works \cite{2022_Arxiv_Cui_dynamic_power_consumption} and \cite{2023_Arxiv_Jin_dynamic_power_consumption}, the authors proposed a comprehensive power consumption model for PIN-diode-based IRSs with respect to factors like polarization, bit resolution, PIN diode states, field programmable gate array (FPGA) control board, drive circuits, etc. In \cite{2023_Dai_PS-DPC}, the authors investigated the energy efficiency maximization problem of an IRS-assisted system with the PS-DPC model. However, they only considered the power constraint at the BS, but do not investigate the cases with a system power budget constraint. In these cases, as we mentioned above, there is a novel power allocation problem to balance the strength of transmitted signals at the BS and the beamforming quality of the IRS.
Furthermore, due to the large number of scattering elements in IRSs, the PS-DPC of IRSs can be quite high. For example, an IRS prototype with 3600 1-bit scattering elements can consume up to $45 \ \mathrm{W}$ \cite{2022_Arxiv_Cui_dynamic_power_consumption}, which is no longer negligible compared to the transmit power at the base station (BS) in a downlink system. In such cases, communication operators may care about the total system power consumption, as they need to pay for the power consumption of both the BS and the IRS. Therefore,  IRS-assisted downlink communication systems with PS-DPC must contend with a total system power constraint that includes the consumption at both the transmitter and the IRS. This system power constraint necessitates a careful balance of the system power distribution between BS and IRS. Specifically, with a fixed system power budget, allocating more power to the BS can enhance the strength of the transmitted signal, but may reduce the power budget available to the IRS. As mentioned above, such a reduction may diminish the passive beamforming DoF at the IRS, adversely impacting the performance of the overall communication system. Therefore, striking an optimal balance in system power allocation between BS and IRS becomes a critical issue for IRS-assisted systems when considering the PS-DPC.
So far, there has been little research on beamforming design for IRS-assisted systems taking into account the PS-DPC. {In \cite{2023_Dai_PS-DPC} and \cite{2024_Jin_PS-DPC}, the authors adopted the PS-DPC model while seeking to maximize the energy efficiency of an IRS-assisted system. However, while these studies focused on maximizing energy efficiency, the rate maximization problem, which is also a critical issue in wireless communications, remains an unsolved challenge in IRS-assisted systems under the PS-DPC model. Additionally, the authors of \cite{2023_Dai_PS-DPC} and \cite{2024_Jin_PS-DPC} primarily focused on the beamforming optimization under a power constraint at the BS, without exploring beamforming strategies under a more comprehensive power budget constraint for the whole system. As mentioned above, such consideration is crucial, particularly in large IRS-assisted systems, where the PS-DPC is a significant factor in the overall system power budget.

In this paper, we investigate the beamforming design for IRS-assisted systems, where we account for the PS-DPC of PIN diodes, aiming at maximizing the achievable rate in both single-user and multi-user cases. To the best of our knowledge, this is the first attempt to tackle the rate maximization problem of IRS-assisted systems following a PS-DPC model. Our main contributions can be summarized as follows:
\begin{itemize}
\item Based on the relationship between the PS-DPC and the IRS coefficients, we formulate a joint power allocation and beamforming optimization problem for IRS-assisted systems under the PS-DPC model. In particular, we maximize the rate of IRS-assisted systems subject to a constraint on the total system power consumption.

\item For the single-user case, we develop a generalized Benders decomposition-based beamforming (GBD-BF) method to tackle the rate maximization problem for the PS-DPC model. This method enables joint optimization of active and passive beamforming at  BS and IRS, respectively, while limiting the system power consumption.

\item Based on statistical channel state information (S-CSI), we further propose a low-complexity beamforming method, where the optimized power allocation strategy between BS and IRS can be optimized offline without frequent updates based on instantaneous CSI. %Based on the optimized power allocation, the beamforming at BS and IRS can then be designed online.

\item In the multi-user case, we transform the sum rate maximization problem into an equivalent weighted mean square error minimization (WMMSE) problem, and propose two novel joint power allocation and beamforming optimization (JPABF) methods to solve it.
%we solve the equivalent weighted mean square error minimization problem via a joint power allocation and phase shift optimization method to optimize the beamforming at both the BS and IRS. Specifically, to tackle the challenging system power constraint introduced by the PS-DPC, we introduce a novel JPABF algorithm that can concurrently optimize the power allocation and phase shift states of IRS elements.

\item We validate the effectiveness of proposed beamforming methods through computer simulations. The results show that our methods can jointly optimize the power allocation and the beamforming at BS and IRS, thereby enhancing the system performance compared to baseline methods. Specifically, the optimized power allocation strategies depend on the available system power budget. When the power budget is high, the IRS can be allocated a sufficient PS-DPC budget to maximize its beamforming quality. However, for a limited power budget, more power is allocated to the BS to enhance the transmit power, resulting in a lower beamforming quality at the IRS due to the reduced PS-DPC budget. %The observed improvements stem from the successful integration of the PS-DPC model into the proposed methods, which allows them to strike an optimized balance between the power consumption at the BS and the IRS.

\end{itemize}

The remainder of this paper is organized as follows. Section \ref{sec:System Model and Problem Formulation} presents the system model and problem formulation for the considered IRS-assisted system with PS-DPC. Section \ref{sec:GBD-BF Algorithm} derives the GBD-BF method for single-user systems. By further exploiting S-CSI, we develop a low-complexity beamforming method in Section \ref{sec:CSC-BF Algorithm: Low-Complexity BF Design Based on Channel Statistical Distribution}. For the multi-user case, Section \ref{sec:PF-WMMSE} proposes two JPABF methods to solve the equivalent WMMSE problem for IRS-assisted systems with PS-DPC. Simulation results are provided in Section \ref{sec:Simulation Results}. Finally, Section \ref{sec:conclusion} presents the conclusions of this paper.

\emph{Notations:} In this paper, $\mathbf{a}$ represents a column vector, with $[\mathbf{a}]_i$ denoting the $i$-th element of $\mathbf{a}$. Similarly, $\mathbf{A}$ represents a matrix, with $[\mathbf{A}]_{ij}$ denoting the $(i,j)$-th element. $(\cdot)^T$, $(\cdot)^H$, $(\cdot)^*$, and $\mathrm{Tr(\cdot)}$ denote the transpose,  conjugate transpose, conjugate, and trace of a matrix, respectively. $|\cdot|$ denotes the modulus of a scalar. $\mathcal{R}\{\cdot\}$ and $\mathcal{I}\{\cdot\}$ denote the real and imaginary parts of a scalar, respectively. $\lfloor \cdot \rfloor $ is the floor function of a real scalar. $\mathrm{mod}_{2}(\cdot)$ denotes the remainder of a real scalar after division by 2. $\mathrm{arcsin}(\cdot)$ and $\mathrm{arccos}(\cdot)$ denote the inverse sine and cosine functions of a real scalar in $[-1,1]$, respectively. $\mathbb{E}(\cdot)$ is the expectation operator. $\circ$ and $\otimes$ denote the Hadamard and Kronecker products, respectively. $\mathrm{diag}(\mathbf{a})$ denotes a diagonal matrix with the elements of $\mathbf{a}$ on its main diagonal, and $\mathrm{diag}(\mathbf{A})$ is a column vector extracting the diagonal elements of $\mathbf{A}$. $\mathrm{blkdiag}(\mathbf{A}_1,\dots,\mathbf{A}_n)$ denotes a block diagonal matrix with diagonal components $\mathbf{A}_1,\dots,\mathbf{A}_n$. $\mathbf{I}_N$ denotes the $N\times N$ identity matrix. $\mathbf{1}_N$ ($\mathbf{0}_N$) denotes the $N \times 1$ all-ones (all-zeros) vector. $\mathbb{R}$ and $\mathbb{R}_0^+$ denote the sets of real and non-negative real numbers, respectively. $\{0,1\}^{N}$ denotes the set of all $N \times 1$ vectors whose elements are restricted to be 1 or 0. $\mathbb{C}^{M \times N}$ represents the set of all ${M \times N}$ complex-valued matrices. $\mathcal{CN}(\mathbf{0}, \mathbf{K})$ denotes the circularly symmetric complex Gaussian distribution with zero mean and covariance matrix $\mathbf{K}$. $\mathcal{U}(a, b)$ denotes the uniform distribution in $[a,b)$.
%$\mathbb{C}^{M \times N}$ denotes the set of all ${M \times N}$ complex-valued matrices. $\mathrm{erf}(x)$ denotes the Gaussian error function. 
% ---------------------------------
% ---- Section II System Model ----
% ---------------------------------

\section{System Model and Problem Formulation}\label{sec:System Model and Problem Formulation}
%\begin{figure}
%	\centering
%	\includegraphics[height=4.6cm,width=7.5cm]{system_model_SU_MISO.eps}
%	\caption{System model of the IRS-assisted downlink single-user MISO system.}\label{fig:system_model_SU_MISO}
%\end{figure}
\begin{figure}
	\centering
	\includegraphics[width=0.4\textwidth]{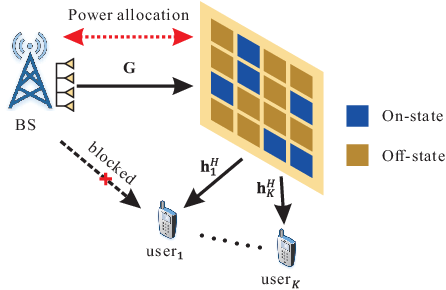}
	\caption{System model for an IRS-assisted downlink multi-user MISO system.}\label{fig:system_model_MU_MISO}
\end{figure}
\subsection{Signal Model}\label{subsec: System Model}
Consider a downlink narrowband multi-user multiple-input single-output (MISO) system as shown in Fig. \ref{fig:system_model_MU_MISO}. The system consists of a BS equipped with an $N$-antenna uniform linear array (ULA) and $K$ single-antenna users. The direct links between the BS and the users are assumed to be blocked, and thus an IRS is deployed to establish virtual line-of-sight (LoS) links to the users. The IRS is designed as a uniform planar array (UPA) of dimension $M_{\mathrm{x}}\times M_{\mathrm{y}}$. The IRS operates with 1-bit phase control resolution, such that the phase shift of each scattering element is binary-switchable between two states through the on/off-state of a PIN diode. Thus, we define ${\mathbf{b}}\in\{0,1\}^{M}$ as the state vector of the employed PIN diodes, where $M=M_{\mathrm{x}}\times M_{\mathrm{y}}$. For the $m$-th element, $[\mathbf{b}]_m=1$ ($[\mathbf{b}]_m=0$) indicates that the corresponding PIN diode is in the on-state (off-state). Without loss of generality, we set the phase shift of the $m$-th element to zero when $[\mathbf{b}]_m=1$ and $\pi$ when $[\mathbf{b}]_m=0$. In this way, the IRS phase shift matrix $\boldsymbol{\Phi}\in\mathbb{C}^{M\times M}$ is given by
\begin{equation}\label{eqn:phi for 1 bit IRS}
\begin{array}{cl}
     \left[\boldsymbol{\Phi}\right]_{mm} &= \left\{\begin{array}{cl}
        -1,&\left[\mathbf{b}\right]_{m}=0,\\
        1, &\left[\mathbf{b}\right]_{m}=1,
    \end{array}\right.   =2\left[\mathbf{b}\right]_{m}-1.\\
\end{array}
\end{equation}
Therefore, the equivalent baseband received signal $y_k$ at the $k$-th user can be represented as
\begin{equation}\label{eqn:baseband received signal}
\begin{array}{cl}
    y_k= \mathbf{h}_k^H(2\mathbf{B}-\mathbf{I}_{M})\mathbf{G}\mathbf{F}\mathbf{s}+n_k,
  %   y& = \mathbf{h}^H\boldsymbol{\Phi}\mathbf{G}\mathbf{f}s+n,  \\
 %    &=\mathbf{h}^H\mathrm{diag}(2\mathbf{b}-\mathbf{1}_{M})\mathbf{G}\mathbf{f}s+n,\\
  %   &{=} (2\mathbf{b}-\mathbf{1}_{M})^T\mathrm{diag}\left(\mathbf{h}^H\right)\mathbf{G}\mathbf{f}s+n,
\end{array}
\end{equation}
where $\mathbf{s}=[s_1,\dots,s_K]^T\in\mathbb{C}^{K\times1}$ is the BS transmitted symbol vector with $\mathbb{E}\{\mathbf{s}\mathbf{s}^H\}=\mathbf{I}_K$, $\mathbf{F}=\big[\mathbf{f}_1,\dots,\mathbf{f}_K\big]\in\mathbb{C}^{N\times K}$ is the active beamforming at the BS, and $\mathbf{B}=\mathrm{diag}(\mathbf{b})$. $n_k\sim\mathcal{CN}(0, \sigma^2)$ represents the additive Gaussian noise at the $k$-th user. $\mathbf{G}\in\mathbb{C}^{M\times N}$ denotes the BS-IRS channel matrix, and $\mathbf{h}_k^H\in\mathbb{C}^{1\times M}$ is the channel vector between the IRS and the $k$-th user.
% The equality $(a)$ follows from the facts that $\mathrm{diag}\big(\boldsymbol{\Phi}\big)=2\mathbf{b}-\mathbf{1}_{M}$, and $\mathbf{p}^H\mathrm{diag}\big(\mathbf{q}\big)=\mathbf{q}^T\mathrm{diag}\big(\mathbf{p}^H\big)$ for arbitrary vectors $\mathbf{p}$ and $\mathbf{q}$. 

\subsection{Channel Model}\label{subsec:Channel Model}
In this paper, we assume the BS-IRS channel $\mathbf{G}$ to be Rician distributed comprising one LoS and a number of non-LoSs (NLoSs) links. Hence, denoting the Rician factor by $\kappa_{\mathrm{G}}$, $\mathbf{G}$ can be modeled as \cite{2022_TWC_IOS}
%In this section, we consider a case where the BS and IRS are deployed at high places with limited scattering, e.g., on the top of buildings. Therefore, the channel between the BS and IRS is dominated by the LoS link, which can be expressed as:
\begin{equation}\label{eqn:channel model for LoS G}
\mathbf{G} = \sqrt{\frac{\kappa_{\mathrm{G}}}{1+\kappa_{\mathrm{G}}}}\mathbf{G}_{\mathrm{LoS}}+\sqrt{\frac{1}{1+\kappa_{\mathrm{G}}}}\mathbf{G}_{\mathrm{NLoS}},
    %\mathbf{G} = \sqrt{MN}\alpha_\mathrm{G} \mathbf{a}_{\mathrm{I}}(\theta_{\mathrm{r}},\varphi_{\mathrm{r}})\mathbf{a}^H_{\mathrm{BS}}(\theta_{\mathrm{t}})+\mathbf{G}_{\mathrm{NLoS}},
\end{equation}
where $\mathbf{G}_{\mathrm{LoS}}\in\mathbb{C}^{M\times N}$ and $\mathbf{G}_{\mathrm{NLoS}}\in\mathbb{C}^{M\times N}$ are the LoS and NLoS components, respectively. Specifically, the LoS component can be characterized by the plane-wave model as $\mathbf{G}_{\mathrm{LoS}} = \sqrt{\mathrm{PL}_\mathrm{G}MN} \mathbf{a}_{\mathrm{I}}(\theta_{\mathrm{r}},\varphi_{\mathrm{r}})\mathbf{a}^H_{\mathrm{BS}}(\theta_{\mathrm{t}})$, 
%\begin{equation}\label{eqn:channel model for LoS G LOS}
%\mathbf{G}_{\mathrm{LoS}} = \sqrt{\mathrm{PL}_\mathrm{G}MN} \mathbf{a}_{\mathrm{I}}(\theta_{\mathrm{r}},\varphi_{\mathrm{r}})\mathbf{a}^H_{\mathrm{BS}}(\theta_{\mathrm{t}}), 
%\end{equation}
where $\mathrm{PL}_\mathrm{G}$, $\theta_{\mathrm{t}}$, $\theta_{\mathrm{r}}$, $\varphi_{\mathrm{r}}$, and $\lambda$ denote the path loss, the angle of departure (AoD), the elevation angle of arrival (AoA), the azimuth AoA of the LoS link, and the carrier wavelength, respectively. The array response vectors for BS and IRS are denoted by $\mathbf{a}_{\mathrm{BS}}$ and $\mathbf{a}_{\mathrm{I}}$, respectively. By defining $\mathbf{a}(N,x)=\frac{1}{\sqrt{N}}\left[1,e^{\mathrm{j}\pi{x}},\dots,e^{\mathrm{j}\pi(N-1){x}}\right]^T$, the array response vector of the ULA at the BS can be expressed as $\mathbf{a}_{\mathrm{BS}}(\theta_{\mathrm{t}}) = \mathbf{a}\big(N,\mathrm{cos}(\theta_{\mathrm{t}})\big)$. For the UPA at the IRS, the array response vector is given by $\mathbf{a}_{\mathrm{I}}(\theta_{\mathrm{r}}, \varphi_{\mathrm{r}}) = \mathbf{a}\big(M_{\mathrm{x}},-\mathrm{sin}(\theta_{\mathrm{r}})\mathrm{sin}(\varphi_{\mathrm{r}})\big)\otimes\mathbf{a}\big(M_{\mathrm{y}},-\mathrm{sin}(\theta_{\mathrm{r}})\mathrm{cos}(\varphi_{\mathrm{r}})\big)$. The NLoS component $\mathbf{G}_{\mathrm{NLoS}}$, on the other hand, can be expressed as $\mathbf{G}_{\mathrm{NLoS}} = \sqrt{\mathrm{PL}_\mathrm{G}}\mathbf{G}_{\mathrm{SS}}, $
%\begin{equation}\label{eqn:channel model for LoS G NLOS}
%$\mathbf{G}_{\mathrm{NLoS}} = \sqrt{\mathrm{PL}_\mathrm{G}}\mathbf{G}_{\mathrm{SS}}, $
%\end{equation}
where the elements of matrix $\mathbf{G}_{\mathrm{SS}}\in\mathbb{C}^{M\times N}$ are complex Gaussian distributed with mean zero and unit covariance, and the distribution of each element is independent of the other elements. %, accounting for the cumulative effect of a large number of scattered links.

Similarly, the IRS-user channel $\mathbf{h}^H_k$ is also modeled as Rician fading, i.e., %\cite{2020_ICC_Rayleigh channel}
\begin{equation}\label{eqn:h gaussian channel}
\mathbf{h}_k^H=\sqrt{\frac{\kappa_{\mathrm{h}_k}}{1+\kappa_{\mathrm{h}_k}}}\mathbf{h}^H_{\mathrm{LoS},k}+\sqrt{\frac{1}{1+\kappa_{\mathrm{h}_k}}}\mathbf{h}^H_{\mathrm{NLoS},k},
\end{equation}
where $\mathbf{h}^H_{\mathrm{LoS},k}=\sqrt{\mathrm{PL}_{\mathrm{h}_k}M} \mathbf{a}^H_{\mathrm{I}}(\vartheta_{\mathrm{t,}k},\psi_{\mathrm{t,}k})$ and $\mathbf{h}_{\mathrm{NLoS},k}\sim\mathcal{CN}(0, \mathrm{PL}_{\mathrm{h}_k}\mathbf{I}_{M})$. $\kappa_{\mathrm{h}_k}$, $\mathrm{PL}_{\mathrm{h}_k}$, $\vartheta_{\mathrm{t,}k}$, and $\psi_{\mathrm{t,}k}$ are the Rician factor, the path loss, the LoS elevation AoD, and the LoS azimuth AoD, respectively. In this paper, we assume that $\mathbf{G}$ and $\mathbf{h}_k$ are perfectly known
at the BS.%\footnote{In this paper, we assume that the channels $\mathbf{G}$ and $\mathbf{h}_k$ are dominated by the LoS link.}, and set $\kappa_{\mathrm{G}}=\kappa_{\mathrm{h}_k}=8$ as in \cite{2003_Rician_factor}.}.

%$\mathbf{h}_{\mathrm{NLoS},k}\sim\mathcal{CN}(\mathbf{0}, \mathrm{PL}_{\mathrm{h}_k}\mathbf{I}_{M})$

\subsection{PS-DPC Model of IRSs}\label{sec:PS-DPC Model}
%The total power consumption of a PIN-diode-based IRS comprises of two components \cite{2022_Arxiv_Cui_dynamic_power_consumption}, \cite{2023_Arxiv_Jin_dynamic_power_consumption}. The first one is the static power consumption, which encompasses the energy used by control devices and circuits. This part of the power consumption is invariant and does not depend on the configuration of the scattering elements of the IRS. The other component is the PS-DPC caused by the PIN diodes, which varies dynamically with respect to the beamforming design at the IRS. Therefore, the total power consumption of the IRS can be modeled as
The total power consumption of a PIN-diode-based IRS comprises two components \cite{2022_Arxiv_Cui_dynamic_power_consumption}, \cite{2023_Arxiv_Jin_dynamic_power_consumption}. The first one is the static power consumption, which encompasses the energy used by control devices and circuits. The second component is the PS-DPC caused by the PIN diodes, which varies dynamically with respect to the beamforming design at the IRS. Therefore, the total power consumption of the IRS can be modeled as
\begin{equation}\label{eqn:IRS power model}
    P_{\mathrm{IRS}} = P_{\mathrm{IRS,static}} +  P_{\mathrm{IRS,PS}},
\end{equation}
where $P_{\mathrm{IRS,static}}$ and $P_{\mathrm{IRS,PS}}$ are the static power consumption and the PS-DPC, respectively. As $P_{\mathrm{IRS,static}}$ is fixed once the IRS is manufactured, in this paper, we mainly focus on the PS-DPC. Specifically, we consider a 1-bit IRS, where the phase shift of each element is switched between two states according to the on/off-states of a PIN diode. Therefore, the PS-DPC of the IRS can be expressed as follows% and its impact on the beamforming design of IRS-assisted systems.
\begin{equation}\label{eqn:dynamic power consumption}
    P_{\mathrm{IRS,PS}} = P_{\mathrm{PIN}}\mathbf{1}_{M }^T{\mathbf{b}},
\end{equation}
where $P_{\mathrm{PIN}}$ is the power consumption of an on-state PIN diode.

%The PS-DPC model results in a strong coupling between the beamforming design and the power consumption of the IRS. Specifically, to provide high-quality beamforming, the IRS may need to turn on a large number of PIN diodes, thus resulting in a high power consumption. However, if the power budget at the IRS is restricted, most of the PIN diodes need to be set to the off-state, which will seriously affect the DoF available for the IRS passive beamforming. Thus, it is necessary to investigate the energy-efficient beamforming design for achieving a good balance between the beamforming quality and the PS-DPC. 

\subsection{System Power Consumption}\label{subsec: System Power Consumption}
In the literature, the transmit power at the BS is typically the only power consumption considered for the wireless system design, regardless of whether \cite{2020_TWC_Zhang_PM_AO} or not \cite{2016_JSAC_Yu_HBF} an IRS is deployed. In contrast, if the PS-DPC of the IRS is also considered, the power consumption of the entire system includes the power consumption of the BS and the IRS, and is given by
\begin{equation}\label{eqn:power consumption model}
\begin{array}{rl}
    % P =& P_\mathrm{BS}+P_\mathrm{IRS} \\
      P=& P_\mathrm{BS,t}+P_\mathrm{BS,C}+N P_\mathrm{BS,RF}\\
      &+P_\mathrm{IRS,static}+P_{\mathrm{PIN}}\mathbf{1}_{M}^T\mathbf{b},
\end{array}
\end{equation}
where $P_\mathrm{BS,t}=\mathrm{Tr}(\mathbf{F}^H\mathbf{F})$ is the BS transmit power consumption. $P_\mathrm{BS,C}$ and $P_\mathrm{BS,RF}$ are the powers consumed by control circuits and each radio frequency chain at the BS, respectively. Note that the power consumption terms in \eqref{eqn:power consumption model} are independent of the beamforming design at BS and IRS, except for $\mathrm{Tr}(\mathbf{F}^H\mathbf{F})$ and $P_{\mathrm{PIN}}\mathbf{1}_{M}^T\mathbf{b}$. Thus, we shall focus on the terms $\mathrm{Tr}(\mathbf{F}^H\mathbf{F})+P_{\mathrm{PIN}}\mathbf{1}_{M}^T\mathbf{b}$ for beamforming design in the following.

\subsection{Problem Formulation}\label{subsec:Problem Formulation}
In this paper, we aim to maximize the sum rate of all users under a system power constraint, which leads to the following optimization problem:
\begin{equation}\label{prb:problem formation multi user}
\begin{array}{cl}
\displaystyle{\maximize_{{{\mathbf{F}}, \mathbf{b}}}} & {\sum_{k=1}^K R_{k} }  \\
\mathrm{subject \; to} & \mathrm{C}1:\mathrm{Tr}(\mathbf{F}^H\mathbf{F})+P_{\mathrm{PIN}}\mathbf{1}_{M}^T\mathbf{b}\leq P_0, \\
& \mathrm{C}2:\mathbf{b}\in\{0,1\}^{M},
\end{array}
\end{equation}
%\begin{equation}\label{eqn:problem formulation}
%\begin{array}{cl}
%\underset{\mathbf{f}, \mathbf{b}}{\operatorname{maximize}} &R\\
%\operatorname { s.t. } & \mathrm{C}1:\mathbf{f}^H\mathbf{f}+P_{\mathrm{PIN}}\mathbf{1}_{M}^T\mathbf{b}\leq P_0, \\
%& \mathrm{C}2:\mathbf{b}\in\{0,1\}^{M},
%\end{array}
%\end{equation}
where $R_k = \mathrm{log}\left(1+\frac{\left|\mathbf{h}_k^H(2\mathbf{B}-\mathbf{I}_{M})\mathbf{G}\mathbf{f}_k\right|^2}{\gamma_k}\right)$ denotes the rate (bps per Hertz) of the $k$-th user,
with ${\gamma}_k=\sum_{k' \neq k}|\mathbf{h}_{k}^H(2\mathbf{B}-\mathbf{I}_{M})\mathbf{G}\mathbf{f}_{k'}|^2+\sigma^2$ representing the interference plus noise. Furthermore, $P_0 = P-P_\mathrm{BS,C}-N P_\mathrm{BS,RF}-P_\mathrm{IRS,static}$ is the effective system power budget.

Due to the binary constraint on $\mathbf{b}$, problem (\ref{prb:problem formation multi user}) is a mixed integer nonlinear programming (MINLP) problem, which is nondeterministic polynomial-time (NP)-hard \cite{2023_Arxiv_Schober_PM_GBD}. Moreover, as we account for the PS-DPC, $\mathbf{F}$ and $\mathbf{b}$ are coupled in both the objective function and the constraint $\mathrm{C1}$. This makes the beamforming optimization more challenging than for conventional IRS-assisted rate maximization problems in the literature, where only the power consumption at the BS is considered \cite{2021_TWC_IRS_MIMO_2}, \cite{2020_TWC_Zhang_PM_AO}, \cite{2020_TWC_Larsson_sum rate maximization}. 

{\emph{Remark 1:} In the literature, classical methods such as the alternating optimization (AO) were adopted to tackle the coupling of $\mathbf{F}$ and $\mathbf{b}$ \cite{2020_TWC_Zhang_PM_AO}. However, this approach cannot be directly applied to solve problem \eqref{prb:problem formation multi user} due to the total system power constraint C1. In particular, if the conventional AO algorithm is adopted to solve problem (\ref{prb:problem formation multi user}), optimizing $\mathbf{F}$ for a fixed $\mathbf{b}$ will always lead to an equality in power constraint $\mathrm{C1}$, i.e., $\mathbf{F}^H\mathbf{F}+P_{\mathrm{PIN}}\mathbf{1}_{M}^T\mathbf{b} = P_0$. However, with this condition, the feasible set during the subsequent optimization of $\mathbf{b}$ is exceedingly small. Specifically, during the $\mathbf{b}$ optimization in the $i$-th iteration, C1 is reformulated as $\mathbf{1}_{M}^T\mathbf{b}\leq \mathbf{1}_{M}^T\mathbf{b}_{(i-1)}$, where $\mathbf{b}_{(i-1)}$ denotes the solution of $\mathbf{b}$ in the last iteration. This means that during the optimization of $\mathbf{b}$, we need to seek a larger achievable rate with \emph{fewer} on-state PIN diodes, which significantly restricts the IRS's beamforming DoF. Therefore, applying the conventional AO algorithm for sum rate maximization under the PS-DPC model results in a quick convergence to highly sub-optimal local maxima. To tackle this challenge, in the following sections, we introduce several novel beamforming methods for both the single-user and multi-user cases, which can jointly optimize the power allocation and the beamforming at both the BS and the IRS. }

\section{GBD-BF Method for Single-User Case}\label{sec:GBD-BF Algorithm}
In this section, we first start with the single-user case, where the BS serves only one user via the IRS. In this case, denoting  $\mathbf{h}^H\in\mathbb{C}^{1\times M}$ and $\mathbf{f}\in\mathbb{C}^{ N\times 1}$ as the IRS-user channel and the active beamforming vector at the BS, respectively, the downlink rate of the user is given by $R=\mathrm{log}\left(1+\frac{\left|(2\mathbf{b}-\mathbf{1}_{M})^T\mathbf{H}_{\mathrm{c}}\mathbf{f}\right|^2}{\sigma^2}\right)$, 
%\begin{equation}\label{eqn:rate SU}
%    \begin{array}{rl}
         %R&=\mathrm{log}\left(1+\frac{\left|\mathbf{h}^H(2\mathbf{B}-\mathbf{I}_{M})\mathbf{G}\mathbf{f}\right|^2}{\sigma^2}\right)   \\
        % &= \mathrm{log}\left(1+\frac{\left|(2\mathbf{b}-\mathbf{1}_{M})^T\mathbf{H}_{\mathrm{c}}\mathbf{f}\right|^2}{\sigma^2}\right),
    %\end{array}
%\end{equation}
where $\mathbf{H}_{\mathrm{c}}=\mathrm{diag}\left(\mathbf{h}^H\right)\mathbf{G}\in\mathbb{C}^{M\times N}$ is the cascaded channel. Hence, the rate maximization problem in \eqref{prb:problem formation multi user} simplifies as follows
\begin{equation}\label{eqn:problem formulation}
\begin{array}{cl}
\underset{\mathbf{f}, \mathbf{b}}{\operatorname{maximize}} &R\\
\operatorname { s.t. } & \mathrm{C}1:\mathbf{f}^H\mathbf{f}+P_{\mathrm{PIN}}\mathbf{1}_{M}^T\mathbf{b}\leq P_0, \\
& \mathrm{C}2:\mathbf{b}\in\{0,1\}^{M}.
\end{array}
\end{equation}
To solve this problem, in this section, we propose a novel method to jointly optimize the beamforming at the BS and the IRS based on the GBD algorithm.

\subsection{Generalized Benders Decomposition }\label{subsec:generalized benders decomposition}
The main idea of the GBD algorithm is to decompose the original MINLP problem into two subproblems, i.e., a primal problem and a master problem, which provide upper and lower bounds for the original problem, respectively \cite{2023_Arxiv_Schober_PM_GBD}, \cite{1972_JOTA_Geoffrion_GBD}, \cite{2023_TWC_Zhu_Han_GBD}. By iteratively solving these two subproblems, the gap between the upper and lower bounds can be progressively reduced, thereby converging towards the optimal solution. To guarantee the convergence of the GBD algorithm, the following two conditions have to be satisfied \cite{2023_Arxiv_Schober_PM_GBD}:
\begin{itemize}
\item The problem is convex with respect to the involved continuous variables if the discrete variables are fixed.
\item The problem is linear with respect to the involved discrete variables if the continuous variables are fixed.
\end{itemize}
To be able to employ the GBD algorithm, we first transform the rate maximization problem in \eqref{eqn:problem formulation}  into an equivalent problem that maximizes the power of the received signal at the user. This transformation leads to the following problem
\begin{equation}\label{eqn:problem formulation2}
\begin{array}{cl}
\underset{\mathbf{f}, \mathbf{b}}{\operatorname{maximize}} &\mathbf{f}^H\mathbf{H}^H_{\mathrm{c}}(2\mathbf{b}-\mathbf{1}_{M})(2\mathbf{b}-\mathbf{1}_{M})^T\mathbf{H}_{\mathrm{c}}\mathbf{f}\\
\operatorname { s.t. } & \mathrm{C}1:\mathbf{f}^H\mathbf{f}+P_{\mathrm{PIN}}\mathbf{1}_{M}^T\mathbf{b}\leq P_0, \\
& \mathrm{C}2:\mathbf{b}\in\{0,1\}^{M}.
\end{array}
\end{equation}
Unfortunately, the objective function of \eqref{eqn:problem formulation2} is still not suitable for applying the GBD algorithm, as it is non-linear with respect to $\mathbf{b}$. However, assuming the optimal solution of \eqref{eqn:problem formulation2} is $\mathbf{f}^\star$, we notice that $\mathrm{e}^{\mathrm{j}\psi}\mathbf{f}^\star$ is also an optimal solution of \eqref{eqn:problem formulation2} for any phase shift $\psi$. This intrinsic property allows us to choose $\psi$ such that the phase of $\mathbf{f}^H\mathbf{H}^H_{\mathrm{c}}(2\mathbf{b}-\mathbf{1}_{M})$ is rotated to zero. Consequently, we can apply this phase rotation to transform \eqref{eqn:problem formulation2} into a more tractable equivalent form:
\begin{equation}\label{eqn:problem formulation3}
\begin{array}{cl}
\underset{\mathbf{f}, \mathbf{b}}{\operatorname{minimize}} &-\mathcal{R}\left\{\mathbf{f}^H\mathbf{H}^H_{\mathrm{c}}(2\mathbf{b}-\mathbf{1}_{M})\right\}\\
\operatorname { s.t. } & \mathrm{C}1:\mathbf{f}^H\mathbf{f}+P_{\mathrm{PIN}}\mathbf{1}_{M}^T\mathbf{b}\leq P_0, \\
& \mathrm{C}2:\mathbf{b}\in\{0,1\}^{M},\\
& \mathrm{C}3:\mathcal{I}\left\{\mathbf{f}^H\mathbf{H}^H_{\mathrm{c}}(2\mathbf{b}-\mathbf{1}_{M})\right\}=0,
\end{array}
\end{equation}
which satisfies the conditions required for application of the GBD algorithm. 

\subsection{Primal Problem: Beamforming Design at the BS}\label{subsec:primal problem}
In the $i$-th iteration, for the primal problem, we fix the discrete variable $\mathbf{b}$ to the value of the last iteration, i.e., $\mathbf{b}_{(i-1)}$. Thus, problem (\ref{eqn:problem formulation3}) reduces to an optimization problem with respect to $\mathbf{f}$, and is given as
\begin{equation}\label{eqn:primal problem}
\begin{array}{cl}
\underset{\mathbf{f}}{\operatorname{minimize}} &-\mathcal{R}\left\{\mathbf{f}^H\mathbf{H}^H_{\mathrm{c}}(2\mathbf{b}_{(i-1)}-\mathbf{1}_{M})\right\}\\
\operatorname { s.t. } & \mathrm{C}1:\mathbf{f}^H\mathbf{f}+P_{\mathrm{PIN}}\mathbf{1}_{M}^T\mathbf{b}_{(i-1)}\leq P_0, \\
& \mathrm{C}3:\mathcal{I}\left\{\mathbf{f}^H\mathbf{H}^H_{\mathrm{c}}(2\mathbf{b}_{(i-1)}-\mathbf{1}_{M})\right\}=0.
\end{array}
\end{equation}
Problem (\ref{eqn:primal problem}) is clearly a convex problem, and the optimal solution, i.e., $\mathbf{f}_{(i)}$, can be computed by standard convex program solvers such as CVX \cite{2020_CVX}. Noticing that $\left(\mathbf{f}_{(i)},\mathbf{b}_{(i-1)}\right)$ is a feasible solution of problem (\ref{eqn:problem formulation3}), the primal problem can provide an upper bound $\eta_{\mathrm{U},(i)}$ for problem \eqref{eqn:problem formulation3}. Furthermore, the Lagrangian function of primal problem (\ref{eqn:primal problem}) can be constructed as
\begin{equation}\label{eqn:primal problem lagrangian function}
\begin{array}{rl}
     L\left(\mathbf{f}, \mathbf{b}_{(i-1)}, \xi, \mu\right)=& -\mathcal{R}\left\{\mathbf{f}^H\mathbf{H}^H_{\mathrm{c}}(2\mathbf{b}_{(i-1)}-\mathbf{1}_{M})\right\}\\
     &+\xi\big(\mathbf{f}^H\mathbf{f}+P_{\mathrm{PIN}}\mathbf{1}_{M}^T\mathbf{b}_{(i-1)}-P_0\big)\\
     
     &\left.+\mu \big(\mathcal{I}\left\{\mathbf{f}^H\right.\mathbf{H}^H_{\mathrm{c}}(2\mathbf{b}_{(i-1)}-\mathbf{1}_{M})\right\}\big),
\end{array}
\end{equation}
where $\xi$ and $\mu$ are the dual variables for constraints $\mathrm{C1}$ and $\mathrm{C3}$, respectively. 

However, problem \eqref{eqn:primal problem} may be infeasible for the given $\mathbf{b}_{(i-1)}$, e.g., when $P_{\mathrm{PIN}}\mathbf{1}_{M}^T\mathbf{b}_{(i-1)}>P_0$. In this case, we consider the following feasibility check problem 
\begin{equation}\label{eqn:primal problem feasiblity}
\begin{array}{cl}
\underset{\mathbf{f}, \delta}{\operatorname{minimize}} &\delta\\
\operatorname { s.t. } & \mathrm{C}1:\mathbf{f}^H\mathbf{f}+P_{\mathrm{PIN}}\mathbf{1}_{M}^T\mathbf{b}_{(i-1)}-P_0\leq \delta, \\
& \mathrm{C}3:\mathcal{I}\left\{\mathbf{f}^H\mathbf{H}^H_{\mathrm{c}}(2\mathbf{b}_{(i-1)}-\mathbf{1}_{M})\right\}=0,
\end{array}
\end{equation}
where $\delta\in\mathbb{R}^+_0$ is an auxiliary variable. By denoting $\overline{\xi}$ and $\overline{\mu}$ as the dual variables for constraints $\mathrm{C1}$ and $\mathrm{C3}$, respectively, we can express the Lagrangian for problem \eqref{eqn:primal problem feasiblity} as follows
\begin{equation}\label{eqn:primal problem feasiblity lagrangian function}
\begin{array}{rl}
     \overline{L}\left(\mathbf{f}, \mathbf{b}_{(i-1)}, \overline{\xi}, \overline{\mu}\right)=&\overline{\xi}\big(\mathbf{f}^H\mathbf{f}+P_{\mathrm{PIN}}\mathbf{1}_{M}^T\mathbf{b}_{(i-1)}-P_0\big)\\
     
     &+\overline{\mu} \big(\mathcal{I}\left\{\mathbf{f}^H\mathbf{H}^H_{\mathrm{c}}(2\mathbf{b}_{(i-1)}-\mathbf{1}_{M})\right\}\big).
\end{array}
\end{equation}
This Lagrangian will be used in the formulation of the master problem, which will be discussed in the next subsection.

\subsection{Master Problem: Design of the IRS}\label{subsec:master problem}
The master problem of the GBD algorithm is obtained based on the nonlinear convex duality theory \cite{1972_JOTA_Geoffrion_GBD}. By respectively denoting $\xi_{(i)}$, $\mu_{(i)}$ and $\overline{\xi}_{(i)}$, $\overline{\mu}_{(i)}$ as the optimal dual solutions of problems (\ref{eqn:primal problem}) and \eqref{eqn:primal problem feasiblity} in the $i$-th iteration, the master problem can be formulated as follows
\begin{equation}\label{eqn:master problem}
\begin{array}{ll}
\underset{\mathbf{b}, \eta_{\mathrm{L},(i)}}{\operatorname{minimize}}  &\eta_{\mathrm{L},(i)} \\
\ \ \ \ \operatorname {s.t.}&\mathrm{C2}:\mathbf{b}\in\{0,1\}^{M},\\
&\mathrm{C4}: \eta_{\mathrm{L},(i)} \geq \underset{\mathbf{f}}{\operatorname{min}}\ L\left(\mathbf{f}, \mathbf{b}, \xi_{(i)}, \mu_{(i)}\right),\forall i \in \mathcal{F}_{(i)},\\
& \mathrm{C5}: 0 \geq \underset{\mathbf{f}}{\operatorname{min}}\ \overline{L}\left(\mathbf{f}, \mathbf{b}, \overline{\xi}_{(i)}, \overline{\mu}_{(i)}\right),\forall i \in \mathcal{I}_{(i)},
\end{array}
\end{equation}
where constraints $\mathrm{C4}$ and $\mathrm{C5}$ represent the optimality and feasibility cuts, respectively. Here, sets $\mathcal{F}{(i)}$ and $\mathcal{I}{(i)}$ contain the iteration indices for which the primal problem was feasible and infeasible, respectively. The optimal value of the master problem, $\eta_{\mathrm{L},(i)}$, establishes a lower bound for problem (\ref{eqn:problem formulation2}) \cite{1972_JOTA_Geoffrion_GBD}. However, it is still challenging to directly solve the master problem in (\ref{eqn:master problem}), because constraints $\mathrm{C4}$ and $\mathrm{C5}$ involve an inner optimization problem related to $\mathbf{f}$. To tackle this issue, we further relax the master problem into the variant 2 (V-2) form of GBD \cite{1972_JOTA_Geoffrion_GBD}. Specifically, constraints $\mathrm{C4}$ and $\mathrm{C5}$ are relaxed by substituting the support functions \big(the right-hand side parts of constraints $\mathrm{C4}$ and $\mathrm{C5}$ in problem \eqref{eqn:master problem}\big) with their local linear approximations at point $\mathbf{f}_{(i)}$. This leads to
\begin{equation}\label{eqn:master problem v2}
    \begin{array}{l}
\underset{\mathbf{b}, \widehat{\eta}_{\mathrm{L},(i)}}{\operatorname{minimize}} \ \widehat{\eta}_{\mathrm{L},(i)} \\
\ \ \ \ \operatorname {s.t.}\ \ \ \mathrm{C2}:\mathbf{b}\in\{0,1\}^{M},\\
\ \ \ \ \ \ \ \ \ \ \ \ \widehat{\mathrm{C4}}:\widehat{\eta}_{\mathrm{L},(i)} \geq L\left(\mathbf{f}_{(i)}, \mathbf{b}, \xi_{(i)}, \mu_{(i)}\right),\forall i \in \mathcal{F}_{(i)},\\

\ \ \ \ \ \ \ \ \ \ \ \ \widehat{\mathrm{C5}}: 0 \geq \ \overline{L}\left(\mathbf{f}_{(i)}, \mathbf{b}, \overline{\xi}_{(i)}, \overline{\mu}_{(i)}\right),\forall i \in \mathcal{I}_{(i)}.
\end{array}
\end{equation}
According to (\ref{eqn:primal problem lagrangian function}) and \eqref{eqn:primal problem feasiblity lagrangian function}, $L$ and $\overline{L}$ are linear with respect to $\mathbf{b}$. Thereby, problem (\ref{eqn:master problem v2}) is a mixed integer linear programming (MILP) problem, which can be directly solved by MILP solvers such as MOSEK \cite{2020_CVX}. The solution obtained, i.e., $\mathbf{b}_{(i)}$, is then employed in the primal problem (\ref{eqn:primal problem}) in the next iteration.

\subsection{GBD-BF Method}\label{subsec:the overall GB-based beamforming scheme}
The overall GBD-BF scheme is summarized in \textbf{Algorithm 1}. In each iteration, we first fix $\mathbf{b}$ to the optimal solution in the last iteration, and solve the primal problem or the feasibility check problem to obtain the optimal beamforming vector $\mathbf{f}$ as well as the optimal dual variables $\xi$ and $\mu$ for the primal problem ($\overline{\xi}$ and $\overline{\mu}$ for the feasibility check problem). Then, the V-2 master problem is solved for fixed $\mathbf{f}$, $\xi$ ($\overline{\xi}$), and $\mu$ ($\overline{\mu}$) to obtain $\mathbf{b}$ for the next iteration. The computational complexity of the GBD-BF method in each iteration is $\mathcal{O}(2^M)$.%, which is mainly attributed to solving the MILP problem \eqref{eqn:master problem v2}.
\begin{algorithm}[t]
\label{alg}
	\caption{GBD-BF method}
	\begin{algorithmic}[1]
        \STATE Set $i$ = 0, $\eta_{\mathrm{U},(0)} = \infty$, $\widehat{\eta}_{\mathrm{L},(0)} = -\infty$, and the tolerance $\Delta$
	\STATE Initialize  $\mathbf{b}_{(0)}$
\REPEAT	
    \STATE $i\leftarrow i+1$
    \IF{the primal problem in (\ref{eqn:primal problem}) is feasible}
	\STATE Solve the primal problem in (\ref{eqn:primal problem})
	\STATE Update $\eta_{\mathrm{U},(i)}$ according to the solution
        %\STATE Obtain $L\left(\mathbf{f}_{(i)}, \mathbf{b}, \xi_{(i)}, \mu_{(i)}\right)$
        \ELSE
        \STATE Solve the feasibility check problem in (\ref{eqn:primal problem feasiblity})
	
        %\STATE Obtain $\overline{L}\left(\mathbf{f}_{(i)}, \mathbf{b}, \overline{\xi}_{(i)}, \overline{\mu}_{(i)}\right)$
        \ENDIF
	\STATE Obtain $\mathbf{b}_{(i)}$ and $\widehat{\eta}_{\mathrm{L},(i)}$ by solving problem (\ref{eqn:master problem v2})
        %\STATE Update $\mathbf{b}_{(i)}$ and $\widehat{\eta}_{\mathrm{L},(i)}$ according to the solution
\UNTIL $\eta_{\mathrm{U},(i)}-\widehat{\eta}_{\mathrm{L},(i)}\leq \Delta$
\STATE Output the solutions $\mathbf{f}^\star=\mathbf{f}_{(i)}$, $\mathbf{b}^\star=\mathbf{b}_{(i)}$
	\end{algorithmic}
\end{algorithm}

\section{Low-Complexity S-CSI-BF Method for Single-User Case}\label{sec:CSC-BF Algorithm: Low-Complexity BF Design Based on Channel Statistical Distribution}
In this section, we propose a low-complexity S-CSI-BF method for IRS-assisted single-user systems with PS-DPC. For this method, we first derive the received SNR as a function of the power consumption at the BS and the IRS exploiting S-CSI, and determine the power allocation strategy offline. Then, the beamforming designs at the BS and the IRS are optimized online based on the obtained power allocation.

\subsection{The Beamforming Design at the BS and the IRS for Given Power Allocation}\label{subsec:BS IRS design}
We first focus on the online beamforming design at the BS and the IRS for given power allocation, i.e., $P_{\mathrm{BS,t}}$ at the BS and $P_{\mathrm{IRS, PS}}$ at the IRS. Specifically, as IRSs are generally deployed such that the channels are dominated by the LoS link \cite{2021_TC_Zhang_IRS_tutorial}, we ignore the NLoS component for algorithm design for simplicity, thus approximating $\mathbf{G}$ and $\mathbf{h}^H$ as%The specific power allocation strategy is further investigated in the next subsection. 

\begin{equation}\label{eqn:channel model approximate}
\begin{array}{l}
     \ \mathbf{G}\approx\mathbf{G}_{\mathrm{LoS}} = {\alpha}_\mathrm{G}\sqrt{MN} \mathbf{a}_{\mathrm{I}}(\theta_{\mathrm{r}},\varphi_{\mathrm{r}})\mathbf{a}^H_{\mathrm{BS}}(\theta_{\mathrm{t}}),  \\
     \mathbf{h}^H\approx\mathbf{h}^H_{\mathrm{LoS}}={\alpha}_\mathrm{h}\sqrt{M} \mathbf{a}^H_{\mathrm{I}}(\vartheta_{\mathrm{t}},\psi_{\mathrm{t}}),
\end{array}
\end{equation}
where ${\alpha}_\mathrm{G}=\sqrt{\frac{\kappa_{\mathrm{G}}}{1+\kappa_{\mathrm{G}}}\mathrm{PL}_{\mathrm{G}}}$ and ${\alpha}_\mathrm{h}=\sqrt{\frac{\kappa_{\mathrm{h}}}{1+\kappa_{\mathrm{h}}}\mathrm{PL}_{\mathrm{h}}}$. In this case, the optimal $\mathbf{f}$ for problem \eqref{eqn:problem formulation} for given $\mathbf{\Phi}$ is obtained as 

\begin{equation}\label{eqn:f with channel model}
    \begin{array}{rl}
         \mathbf{f}&=  \epsilon \mathbf{G}^H\boldsymbol{\Phi}^H\mathbf{h}\\
         & \approx\epsilon {\alpha}^*_{\mathrm{G}}{\alpha}^*_{\mathrm{h}}M\sqrt{N}\mathbf{a}_{\mathrm{BS}}(\theta_{\mathrm{t}})\mathbf{a}^H_{\mathrm{I}}(\theta_{\mathrm{r}},\varphi_{\mathrm{r}})\boldsymbol{\Phi}^H\mathbf{a}_{\mathrm{I}}(\vartheta_{\mathrm{t}},\psi_{\mathrm{t}}),
    \end{array}
\end{equation}
where $\epsilon=\sqrt{\frac{{P_{\mathrm{BS,t}}}}{\mathbf{h}^H\boldsymbol{\Phi}\mathbf{G}\mathbf{G}^H\boldsymbol{\Phi}^H\mathbf{h}}}$. Therefore, the received SNR at the user can be expressed as:
\begin{equation}\label{eqn:received SNR with channel model}
    \begin{array}{rl}
         \mathrm{SNR} & = \frac{1}{\sigma^2}\mathbf{f}^H\mathbf{G}^H\boldsymbol{\Phi}^H\mathbf{h}\mathbf{h}^H\boldsymbol{\Phi}\mathbf{G}\mathbf{f}  \\
         %& =\frac{1}{\sigma^2}\epsilon^2\mathbf{h}^H\boldsymbol{\Phi}\mathbf{G}\mathbf{G}^H\boldsymbol{\Phi}^H\mathbf{h}\mathbf{h}^H\boldsymbol{\Phi}\mathbf{G}\mathbf{G}^H\boldsymbol{\Phi}^H\mathbf{h}\\
         %& \approx\frac{1}{\sigma^2}\epsilon^2|{\alpha}_\mathrm{G}|^4 M^2N^2\mathbf{h}^H\boldsymbol{\Phi}\mathbf{a}_{\mathrm{I}}(\theta_{\mathrm{r}},\varphi_{\mathrm{r}})\mathbf{a}^H_{\mathrm{I}}(\theta_{\mathrm{r}},\varphi_{\mathrm{r}})\boldsymbol{\Phi}^H\mathbf{h}\\
         %&\ \ \ \times\mathbf{h}^H\boldsymbol{\Phi}\mathbf{a}_{\mathrm{I}}(\theta_{\mathrm{r}},\varphi_{\mathrm{r}})\mathbf{a}_{\mathrm{I}}^H(\theta_{\mathrm{r}},\varphi_{\mathrm{r}})\boldsymbol{\Phi}^H\mathbf{h}\\
        % & =\frac{1}{\sigma^2}P_{\mathrm{BS,t}}|\alpha_\mathrm{G}|^2 MN\mathbf{h}^H\boldsymbol{\Phi}\mathbf{a}_{\mathrm{I}}(\theta_{\mathrm{r}},\varphi_{\mathrm{r}})\mathbf{a}^H_{\mathrm{I}}(\theta_{\mathrm{r}},\varphi_{\mathrm{r}})\boldsymbol{\Phi}^H\mathbf{h}\\
         & \approx\frac{P_{\mathrm{BS,t}}}{\sigma^2}|{\alpha}_\mathrm{G}{\alpha}_\mathrm{h}|^2 M^2N\left| \mathbf{a}^H_{\mathrm{I}}(\vartheta_{\mathrm{t}},\psi_{\mathrm{t}})\boldsymbol{\Phi}\mathbf{a}_{\mathrm{I}}(\theta_{\mathrm{r}},\varphi_{\mathrm{r}})\right|^2\\
         & =\frac{P_{\mathrm{BS,t}}}{\sigma^2}|{\alpha}_\mathrm{G}{\alpha}_\mathrm{h}|^2 M^2N\left|\mathbf{h}_{\mathrm{o}}^H\boldsymbol{\phi}\right|^2,\\
    \end{array}
\end{equation}
where $\mathbf{h}_{\mathrm{o}}^H =  \mathbf{a}^H_{\mathrm{I}}(\vartheta_{\mathrm{t}},\psi_{\mathrm{t}}) \circ \mathbf{a}^T_{\mathrm{I}}(\theta_{\mathrm{r}},\varphi_{\mathrm{r}})$ and $\boldsymbol{\phi}=\mathrm{diag}(\mathbf{\Phi})$. To maximize the SNR, the continuous phase shift of the $m$-th IRS element without considering the PS-DPC is given by 
\begin{equation}\label{eqn:continue phi}
    [\boldsymbol{\widehat{\phi}}^\star]_m ={[\mathbf{h}_{\mathrm{o}}]_m}\big/{\big|[\mathbf{h}_{\mathrm{o}}]_m\big|}.
\end{equation}
However, for a 1-bit IRS, the phase shifts of the IRS elements are constrained to binary states, i.e., $0$ or $\pi$. Moreover, due to the PS-DPC, the number of elements with phase shift $\pi$ (${M}_{\mathrm{on}}$) is limited by the available power budget for the IRS, i.e., ${M}_{\mathrm{on}}=  \lfloor \frac{P_{\mathrm{IRS, PS}}}{P_{\mathrm{PIN}}} \rfloor $. In this case, {\color{black}by rounding \eqref{eqn:continue phi} to the binary phase shift states and restricting ${M}_{\mathrm{on}}\leq \lfloor \frac{P_{\mathrm{IRS, PS}}}{P_{\mathrm{PIN}}} \rfloor $}, the practical phase shift configuration of the $m$-th scattering element is given by 
\begin{equation}\label{eqn:BF of CSC-BF}
    [\boldsymbol{{\phi}}^\star]_m=\left\{\begin{array}{cc}
         1,&m \in \mathcal{M},  \\
         -1,& \mathrm{otherwise},
    \end{array}\right.
\end{equation}
where the set $\mathcal{M}$ is defined by the following process: First, we sort the elements of $\mathbf{h}^H_\mathrm{o}$ in descending order based on their real parts. Then, we select the indices of the top $\widetilde{M}_\mathrm{on}$ elements from this sorted list to form the set $\mathcal{M}$. Here, $\widetilde{M}_\mathrm{on}$ is defined as $\mathrm{min}(M_\mathrm{p}, M_{\mathrm{on}})$, where $M_\mathrm{p}$ represents the number of positive elements in $\mathcal{R}\big\{{\mathbf{h}_{\mathrm{o}}}\big\}$.

\subsection{The Power Allocation Design}\label{subsec: Obtaining the optimal on/off ratio of IRS elements}
Based on the online beamforming design presented in Section \ref{subsec:BS IRS design}, in this subsection, we reformulate the received SNR as a function of the power consumption at the BS and the IRS by exploiting S-CSI. Based on this relationship, the power allocation design that satisfies the system power constraint can be optimized offline without frequent updates based on instantaneous CSI. 

%In this subsection, we optimize the power allocation between the BS and IRS, i.e., $P_{\mathrm{BS,t}}$ and ${\widetilde{M}_\mathrm{on}}$, based on beamforming design presented in the last subsection. Specifically, our focus is on optimizing ${\widetilde{M}_\mathrm{on}}$, as the transmit signal power at the BS can be determined by $P_{\mathrm{BS,t}}=P_{0}-\widetilde{M}_{\mathrm{on}}P_{\mathrm{PIN}}$. 
Specifically, according to the channel model in Section \ref{subsec:Channel Model}, we can further rewrite $\mathbf{h}_{\mathrm{o}}^H$ as follows
\begin{equation}\label{eqn:ho}
    \begin{array}{rl}
        \mathbf{h}_{\mathrm{o}}^H &= \mathbf{a}^H_{\mathrm{I}}(\vartheta_{\mathrm{t}},\psi_{\mathrm{t}}) \circ \mathbf{a}^T_{\mathrm{I}}(\theta_{\mathrm{r}},\varphi_{\mathrm{r}})   \\
         & = \frac{1}{\sqrt{M}}\mathbf{a}^H(M_{\mathrm{x}}, \widehat{\theta})\otimes \mathbf{a}^H(M_{\mathrm{y}}, \widehat{\varphi}),
    \end{array}
\end{equation}
where $\widehat{\theta}=-\sin{(\vartheta_{\mathrm{t}})}\sin{(\psi_{\mathrm{t}})}+\sin{(\theta_{\mathrm{r}})}\sin{(\varphi_{\mathrm{r}})}$ and $\widehat{\varphi}=-\sin{(\vartheta_{\mathrm{t}})}\cos{(\psi_{\mathrm{t}})}+\sin{(\theta_{\mathrm{r}})}\cos{(\varphi_{\mathrm{r}})}$. In this case, the elements in $\mathbf{h}_{\mathrm{o}}^H$ have the following property.
\begin{lemma}
    If $\widehat{\theta}$ or $\widehat{\varphi}$ is an irrational number, when $M_\mathrm{x}\rightarrow\infty$ and $M_\mathrm{y}\rightarrow\infty$, the sequence of all elements in $\mathbf{h}_{\mathrm{o}}^H$ approaches a distribution described by random variable $h_{\mathrm{o}}$ as\footnote{Although Lemma 1 is rigorous only when $\widehat{\theta}$ or $\widehat{\varphi}$ is an irrational number, our simulations in Section \ref{sec:Simulation Results} shall verify that the proposed S-CSI-BF method is still effective in scenarios where both $\widehat{\theta}$ and $\widehat{\varphi}$ are rational numbers. }
\begin{equation}\label{eqn:ho E}
    h_{\mathrm{o}}=\frac{1}{M}e^{\mathrm{j}v},
\end{equation}
where $v\sim\mathcal{U}(0, 2\pi)$.
\end{lemma} 
\emph{Proof}: 
Without loss of generality, we assume $\widehat{\varphi}$ to be an irrational number. In this case, according to the Weyl's Equidistribution Theorem \cite{Uniform Distribution of Sequences}, the sequence $\mathbf{v}^H_{[x]}=[\mathrm{mod}_{2}(-x), \mathrm{mod}_{2}\big(-x-\widehat{\varphi}), \dots,\mathrm{mod}_{2}(-x-(M_{\mathrm{y}}-1)\widehat{\varphi}\big)]$ is uniformly distributed in $[0,2)$ for any $x\in\mathbb{R}$ when $M_{\mathrm{y}}\rightarrow\infty$. Furthermore, note that $\mathbf{h}_\mathrm{o}$ in \eqref{eqn:ho} can be represented by $\mathbf{v}^H_{[x]}$ as
\begin{equation}\label{eqn:ho2}
\begin{array}{l}
     \mathbf{h}_{\mathrm{o}}^H = \frac{1}{\sqrt{M M_{\mathrm{x}}}}\Big[\mathbf{a}^H(M_{\mathrm{y}}, \widehat{\varphi}), \ e^{-\mathrm{j}\pi\widehat{\theta}}\mathbf{a}^H(M_{\mathrm{y}}, \widehat{\varphi}),\dots,\\
       \ \ \ \ \ \ \ \ \ \ \ \ \ \ \ \ e^{-\mathrm{j}\pi(M_{\mathrm{x}}-1)\widehat{\theta}}\mathbf{a}^H(M_{\mathrm{y}}, \widehat{\varphi})\Big]\\
     %\ \ \ \ = \frac{1}{M}\Big[e^{\mathrm{j}\pi\mathbf{v}_{[0]}^H}, e^{-\mathrm{j}\pi\widehat{\theta}}e^{\mathrm{j}\pi\mathbf{v}_{[0]}^H},\dots, e^{-\mathrm{j}\pi(M_{\mathrm{x}}-1)\widehat{\theta}}e^{\mathrm{j}\pi\mathbf{v}_{[0]}^H}\Big]\\
     \ \ \ \ = \frac{1}{M}\Big[e^{\mathrm{j}\pi\mathbf{v}_{[0]}^H}, e^{\mathrm{j}\pi\mathbf{v}_{[\widehat{\theta}]}^H},\dots,e^{\mathrm{j}\pi\mathbf{v}_{[(M_{\mathrm{x}}-1)\widehat{\theta}]}^H}\Big].
\end{array}
\end{equation}
As $\mathbf{v}_{[0]}^H,\ \mathbf{v}_{[\widehat{\theta}]}^H,\ \dots,\ \mathbf{v}_{[(M_{\mathrm{x}}-1)\widehat{\theta}]}^H$ are uniformly distributed in $[0,2)$, we can prove that the sequence of all elements of $\mathbf{h}_{\mathrm{o}}^H$ approaches to the distribution described by the random variable $h_{\mathrm{o}}$ in \eqref{eqn:ho E}, which completes the proof.$\hfill\blacksquare$

{\color{black}Therefore, by assuming $M_\mathrm{x}\rightarrow\infty$ and $M_\mathrm{y}\rightarrow\infty$, we can replace  $\mathbf{h}_{\mathrm{o}}^H\boldsymbol{\phi}^\star$
by an expectation as follows}\footnote{{\color{black}Although the number of IRS elements in practical systems is always finite, assuming an infinite number of elements is an efficient asymptotic method to simplify analysis and provide valuable insights. Hence, this approach has been widely applied in previous works \cite{2020_TWC_Zhang_PM_AO}, \cite{STAR_IRS_phase_couple_2021}.}}
\begin{equation}\label{eqn:transformation of imag 1}
    \begin{array}{rl}
        \mathbf{h}_{\mathrm{o}}^H\boldsymbol{\phi}^\star=&\sum_{m=1}^M[\mathbf{h}^*_{\mathrm{o}}]_m[\boldsymbol{\phi}^\star]_m \\ 
         \approx&M\mathbb{E}\left\{h_{\mathrm{o}}{\varphi}^\star\right\},
    \end{array}
\end{equation}
where  
\begin{equation}\label{eqn:phi design}
    \varphi^\star=\left\{\begin{array}{cc}
         1,& \mathcal{R}\left\{ h_{\mathrm{o}}\right\}>\tau, \\
         -1,& \mathcal{R}\left\{ h_{\mathrm{o}}\right\}\leq\tau,
    \end{array}\right.
\end{equation}
with $\tau\in[0, \frac{1}{M})$ denoting the value corresponding to $P\left(\mathcal{R}\left\{ h_{\mathrm{o}}\right\}>\tau\right)=\frac{\widetilde{M}_\mathrm{on}}{M}$. Specifically, the probability density function (PDF) of $\mathcal{R}\left\{{h}_{\mathrm{o}}\right\}$ is given by
\begin{equation}\label{eqn:pdf of h}
    \mathcal{P}_{\mathcal{R}\left\{{h}_{\mathrm{o}}\right\}}(x)=\frac{M}{\pi\sqrt{1-M^2x^2}}.
\end{equation}
By further defining $t =\mathrm{arcsin}(M\tau)\in[0,\frac{\pi}{2})$, the probability of $\mathcal{R}\left\{ h_{\mathrm{o}}\right\}>\tau$ can be formulated as
\begin{equation}\label{eqn:probability of tau}
\begin{array}{c}
     P\left(\mathcal{R}\left\{ h_{\mathrm{o}}\right\}>\tau\right)=\int_{\tau}^{\frac{1}{M}} \frac{M}{\pi\sqrt{1-M^2x^2}} \mathrm{d}x=\frac{1}{2}-\frac{1}{\pi}t, \\
\end{array}
\end{equation}
which implies that $\frac{\widetilde{M}_\mathrm{on}}{M}=\frac{1}{2}-\frac{1}{\pi}t$. In this case, we can denote the power consumption at IRS and BS as a function of $t$:
\begin{equation}\label{eqn:power as function of t}
    \begin{array}{l}
         P_{\mathrm{IRS, PS}}= P_{\mathrm{PIN}}\widetilde{M}_\mathrm{on}=P_{\mathrm{PIN}}(\frac{M}{2}-\frac{M}{\pi}t),  \\
         P_{\mathrm{BS,t}}=P_0-P_{\mathrm{IRS, PS}} = P_0-P_{\mathrm{PIN}}(\frac{M}{2}-\frac{M}{\pi}t).
    \end{array}
\end{equation}
%Note that in \eqref{eqn:power as function of t}, the system power constraint is still satisfied, i.e., $P_{\mathrm{IRS, PS}}+P_{\mathrm{BS,t}}=P_0$. 
Therefore, the essential part of power allocation optimization is to express the SNR as a function of $t$, and then find the optimal $t$ that maximizes the SNR. To achieve this, we further establish two key properties of $h_{\mathrm{o}}\varphi^\star$ in the following lemmas.
\begin{lemma}\label{lem:imag of ho=0}
$\mathbb{E}\left\{\mathcal{I}\left\{{h}_{\mathrm{o}}{\varphi}^\star\right\}\right\}=0.$
\end{lemma}
\emph{Proof}: 
According to \eqref{eqn:ho E}, we have $\mathcal{R}\left\{{h}_{\mathrm{o}}\right\}=\frac{1}{M}\cos{(v)}$ and $\mathcal{I}\left\{{h}_{\mathrm{o}}\right\}=\frac{1}{M}\sin{(v)}$. In this case, if $\mathcal{R}\left\{{h}_{\mathrm{o}}\right\}>\tau$, it means $v\in[0,\widehat{\tau})\cup(2\pi-\widehat{\tau},2\pi)$ with $\widehat{\tau}=\mathrm{arccos}(M\tau)$. Therefore, we can further rewrite $\mathcal{I}\left\{{h}_{\mathrm{o}}{\varphi}^\star\right\}$ as
\begin{equation}\label{eqn:expectation v1phi1 imagnary}
    \mathcal{I}\left\{{h}_{\mathrm{o}}{\varphi}^\star\right\}=\left\{\begin{array}{l}
         \ \frac{1}{{M_1}}\sin(v),\ v\in[0,\widehat{\tau})\cup(2\pi-\widehat{\tau},2\pi),\\
         -\frac{1}{{M_1}}\sin(v),v\in[\widehat{\tau}, 2\pi-\widehat{\tau}],
    \end{array}\right.
\end{equation}
which means $\mathbb{E}\left\{\mathcal{I}\left\{{h}_{\mathrm{o}}{\varphi}^\star\right\}\right\}=0$ as $v\sim\mathcal{U}(0, 2\pi)$.$\hfill\blacksquare$
\begin{lemma}\label{lem:E of h phi}
$\mathbb{E}\left\{\mathcal{R}\left\{{h}_{\mathrm{o}}{\varphi}^\star\right\}\right\}=\frac{2}{M\pi}\cos{t}.$
\end{lemma}
\emph{Proof}: According to \eqref{eqn:phi design} and \eqref{eqn:pdf of h}, the PDF of $\mathcal{R}\left\{{h}_{\mathrm{o}}\right\}\varphi^\star$ can be written as 
\begin{equation}
    \mathcal{P}_{\mathcal{R}\left\{{h}_{\mathrm{o}}\varphi^\star\right\}}(x)=\left\{\begin{array}{cc}
         0,&  -\frac{1}{M}\leq x<-\tau,\\
         \frac{M}{\pi\sqrt{1-M^2x^2}},& -\tau\leq x\leq \tau, \\
         \frac{2M}{\pi\sqrt{1-M^2x^2}}, &\tau<x\leq \frac{1}{M},
    \end{array}\right.
\end{equation}
therefore, $\mathbb{E}\left\{\mathcal{R}\left\{{h}_{\mathrm{o}}{\varphi}^\star\right\}\right\}$ can be calculated by
\begin{equation}
   \begin{array}{rl}
        \mathbb{E}\left\{\mathcal{R}\left\{{h}_{\mathrm{o}}{\varphi}^\star\right\}\right\}&=\int_{-\frac{1}{M}}^{\frac{1}{M}}x\mathcal{P}_{\mathcal{R}\left\{{h}_{\mathrm{o}}\varphi^\star\right\}}(x)\mathrm{d}x  \\
        & =\int_{\tau}^{\frac{1}{M}}\frac{2Mx}{\pi\sqrt{1-M^2x^2}}\mathrm{d}x\\
        %& =\frac{2}{M\pi}\sqrt{1-M^2\tau^2} \\
        & =\frac{2}{M\pi}\cos{t}, \\

   \end{array}
\end{equation}
which completes the proof of Lemma 3.$\hfill\blacksquare$

{\color{black}By exploiting the properties in Lemma 2 and Lemma 3, we ignore the term $\mathbb{E}\left\{\mathcal{I}\left\{{h}_{\mathrm{o}}{\varphi}^\star\right\}\right\}$, and further reformulate the SNR as a function of $t$, which leads to}
\begin{equation}\label{eqn:received SNR with channel model3}
    \begin{array}{rl}
         \mathrm{SNR} & \approx\frac{P_{\mathrm{BS,t}}}{\sigma^2}|{\alpha}_\mathrm{G}{\alpha}_\mathrm{h}|^2 M^2N\left|\mathbf{h}_{\mathrm{o}}^H\boldsymbol{\phi}\right|^2\\
         & \approx \frac{P_{\mathrm{BS,t}}}{\sigma^2}|{\alpha}_\mathrm{G}{\alpha}_\mathrm{h}|^2 M^4N\left|\mathbb{E}\left\{\mathcal{R}\left\{{h}_{\mathrm{o}}{\varphi}^\star\right\}\right\}\right|^2\\
         & = \frac{4|{\alpha}_\mathrm{G}{\alpha}_\mathrm{h}|^2M^2N}{\pi^2\sigma^2}\left(P_0-\frac{P_{\mathrm{PIN}}M}{2}+\frac{P_{\mathrm{PIN}}M}{\pi}t\right) \mathrm{cos}^2t,
    \end{array}
\end{equation}
the maximum of which is given in the following Lemma.

\begin{lemma}
    The maximum of \eqref{eqn:received SNR with channel model3} can be found by solving the following equation
    \begin{equation}\label{eqn:derivation equal to zero}
    \left(\frac{2\pi P_0}{P_{\mathrm{PIN}}M}-\pi+2t^\star\right)^{-1}=\tan{t^\star}.
\end{equation}
\end{lemma}
\emph{Proof}: To find the maximum value of the SNR in \eqref{eqn:received SNR with channel model3}, we first calculate the derivative of the SNR with respect to $t$ as
\begin{equation}\label{eqn:SNR derivative}
    \begin{array}{rl}
         \frac{\mathrm{d}\big(\mathrm{SNR}\big)}{\mathrm{d}t}
         & = \frac{4|{\alpha}_\mathrm{G}{\alpha}_\mathrm{h}|^2M^2N}{\pi^2\sigma^2}\cos^2{t}\Big[\frac{P_{\mathrm{PIN}}M}{\pi}-2\tan{t}\\
         &\ \ \ \times\left(P_0-\frac{P_{\mathrm{PIN}}M}{2}+\frac{P_{\mathrm{PIN}}M}{\pi}t\right)\Big].
    \end{array}
\end{equation}
As $\frac{\mathrm{d}\big(\mathrm{SNR}\big)}{\mathrm{d}t}\Big|_{t=0}= \frac{4|{\alpha}_\mathrm{G}{\alpha}_\mathrm{h}|^2M^2N}{\pi^2\sigma^2}\frac{P_{\mathrm{PIN}}M}{\pi}>0$ and $\frac{\mathrm{d}\big(\mathrm{SNR}\big)}{\mathrm{d}t}\Big|_{t\rightarrow\frac{\pi}{2}}=\frac{4|{\alpha}_\mathrm{G}{\alpha}_\mathrm{h}|^2M^2N}{\pi^2\sigma^2}(-2P_0\sin{t}\cos{t})$ tends to zero from below, the derivative in \eqref{eqn:SNR derivative} must have at least one zero point, which can be calculated from
\begin{equation}\label{eqn:zero point of SNR derivative}
   \left(\frac{2\pi P_0}{P_{\mathrm{PIN}}M}-\pi+2t\right)^{-1}=\tan{t}.
\end{equation}
For $t\in[0,\frac{\pi}{2})$ and $\left(\frac{2\pi P_0}{P_{\mathrm{PIN}}M}-\pi+2t\right)^{-1}>0$, the left-hand side of \eqref{eqn:zero point of SNR derivative} is a decreasing function, while the right-hand side is an increasing function. Therefore, there can only be one zero point of \eqref{eqn:zero point of SNR derivative}, which is the maximum point of \eqref{eqn:received SNR with channel model3}, i.e., $t^\star$.$\hfill\blacksquare$

Once $t^\star$ is found according to Lemma 4, the power allocation strategy that satisfies the system power constraint can be determined based on \eqref{eqn:power as function of t}. Since \eqref{eqn:derivation equal to zero} does not depend on instantaneous CSI, the power allocation can be optimized offline without frequent updates.%, and finally obtain the optimal power allocation ratio.
%\begin{algorithm}[t]
%\label{alg}
%	\caption{The CSC-BF method}
%	\begin{algorithmic}[1]
 %       \STATE Calculate the optimal $t^\star$ for given $P_0$, $P_{\mathrm{PIN}}$, and $M$ according to \eqref{eqn:derivation equal to zero}.
%        \STATE Calculate the number of on-state PIN diodes at the IRS $\widetilde{M}_\mathrm{on}=\frac{1}{2}M-\frac{1}{2}M\mathrm{erf}\left(t^\star\right)$ and the signal power $P_{\mathrm{BS,t}}=P_{0}-\widetilde{M}_{\mathrm{on}}P_{\mathrm{PIN}}$ at the BS.
 %       \STATE Calculate the optimal $\mathbf{f}^\star$ and $\boldsymbol{\phi}^\star$ according to \eqref{eqn:f with channel model} and \eqref{eqn:BF of CSC-BF}, respectively.
%\STATE Output the optimal $\mathbf{f}^\star$, $\boldsymbol{\phi}^\star$.
%	\end{algorithmic}
%\end{algorithm}

\subsection{S-CSI-BF Method}\label{subsec:CSC-BF Algorithm}
The S-CSI-BF method is summarized as follows. In the first step, we obtain the $t^\star$ that maximizes the received SNR by solving  \eqref{eqn:derivation equal to zero}, thus determining the power allocation between BS and IRS offline. With the determined power allocation, the beamforming at the BS and the IRS can then be respectively optimized according to \eqref{eqn:f with channel model} and \eqref{eqn:BF of CSC-BF}. The complexity of the S-CSI-BF method is $\mathcal{O}\left(MN+M\mathrm{log}M\right)$, which is mainly attributed to the calculation of $\mathbf{f}$ and the sorting of $\mathcal{R}\big\{\mathbf{h}_\mathrm{o}\big\}$. 

{\color{black}It should be noted that during the derivation of the S-CSI-BF method, certain approximations were applied, e.g., ignoring the NLoS components of channels \big(in \eqref{eqn:channel model approximate}\big), assuming $M_\mathrm{x}\rightarrow\infty$ and $M_\mathrm{y}\rightarrow\infty$ \big(in \eqref{eqn:transformation of imag 1}\big), and ignoring the term $\mathbb{E}\left\{\mathcal{I}\left\{{h}_{\mathrm{o}}{\varphi}^\star\right\}\right\}$ \big(in \eqref{eqn:received SNR with channel model3}\big). }Despite these simplifications, our numerical results in Section \ref{sec:Simulation Results} shall verify the effectiveness of the S-CSI-BF method. More importantly, the computational complexity of the S-CSI-BF method is much lower than that of the GBD-BF method, as it avoids the high-complexity iteration processes and the MILP master problem in \eqref{eqn:master problem v2}.

\section{JPABF Methods for Multi-User Case}\label{sec:PF-WMMSE}
{In this section, we further consider the general multi-user case, and reformulate the sum rate maximization problem as an equivalent WMMSE problem. To effectively handle the power allocation challenge introduced by the PS-DPC model, we propose two novel JPABF methods, which enable the joint optimization of the power allocation and the beamforming at the BS and the IRS under the system power constraint.

\subsection{WMMSE Approach}\label{subsec:System model multi-user}
By referring to \cite{2021_Zhu_WMMSE_HBF}, we can equivalently transform the sum rate maximization problem \eqref{prb:problem formation multi user} into the following WMMSE problem
\begin{equation}\label{prb:WMMSE problem multi user}
\begin{array}{cl}
\displaystyle{\minimize_{{\zeta, {\widehat{\mathbf{F}}}, \mathbf{b}, {w}_k, {\psi}_k}}} & {g=\sum_{k=1}^K ({\psi}_k{e}_k-\mathrm{ln}\ {\psi}_k }) \\
\mathrm{subject \; to} & \mathrm{C}1:\zeta^2\mathrm{Tr}(\widehat{\mathbf{F}}^H\widehat{\mathbf{F}})+P_{\mathrm{PIN}}\mathbf{1}_{M}^T\mathbf{b}\leq P_0, \\
& \mathrm{C}2:\mathbf{b}\in\{0,1\}^{M},
\end{array}
\end{equation}
where $\mathbf{F} = \zeta \widehat{\mathbf{F}}$, with $\zeta\in\mathbb{R}_0^+$ being an auxiliary variable and $\widehat{\mathbf{F}}\in \mathbb{C}^{N\times K}$. ${e}_k=\mathbb{E}[|\zeta^{-1}{w}^*_k{y}_k-{s}_k|^2]$ denotes the MSE of the $k$-th user, with ${w}_k$, ${\psi}_k \in\mathbb{C}$ being auxiliary variables. Unfortunately, problem \eqref{prb:WMMSE problem multi user} is intractable due to the coupling of multiple optimization variables and the highly
non-convex binary constraint on $\mathbf{b}$. Hence, a globally optimal solution cannot be obtained in general. To tackle this problem, we apply the AO principle, and alternately optimize the auxiliary variables (${w}_{k}$ and ${\psi}_{k}$) and the beamforming variables ($\zeta$, $\widehat{{\mathbf{F}}}$, and ${\mathbf{b}}$) to achieve a locally optimal solution of the WMMSE problem. Specifically, by denoting $\mathbf{h}_{\mathrm{e},k}^H=\mathbf{h}_k^H(2\mathbf{B}+\mathbf{I}_M)\mathbf{G}\in\mathbb{C}^{1\times N}$, the solutions of ${w}_{k}$ and ${\psi}_{k}$ can be given by

\begin{equation}\label{eqn:W psi}
    w_{k}=\zeta\left(\mathbf{h}_{\mathrm{e}, k}^H \mathbf{f}_k \mathbf{f}_k^H \mathbf{h}_{\mathrm{e}, k}+\gamma_k\right)^{-1} \mathbf{h}_{\mathrm{e}, k}^H \mathbf{f}_k, \quad \psi_k=e_k^{-1}.
\end{equation}
With ${w}_{k}$ and ${\psi}_{k}$ determined, problem \eqref{prb:WMMSE problem multi user} can then be redefined as the following optimization problem with respect to $\zeta$, $\widehat{\mathbf{F}}$, and $\mathbf{b}$, excluding constant terms
\begin{equation}\label{prb:WMMSE problem multi user1}
\begin{array}{cl}
\displaystyle{\minimize_{\zeta, \widehat{\mathbf{F}}, \mathbf{b}}} & \overline{g} \\
\mathrm{subject \; to} & \mathrm{C}1:\zeta^2\mathrm{Tr}(\widehat{\mathbf{F}}^H\widehat{\mathbf{F}})+P_{\mathrm{PIN}}\mathbf{1}_{M}^T\mathbf{b}\leq P_0, \\
& \mathrm{C}2:\mathbf{b}\in\{0,1\}^{M},
\end{array}
\end{equation}
where $\overline{g}=\mathrm{Tr}(\mathbf{\Psi}+\mathbf{\Psi}{\mathbf{W}}^H\mathbf{H}_{\mathrm{e}}\widehat{\mathbf{F}}\widehat{\mathbf{F}}^H\mathbf{H}_{\mathrm{e}}^H\mathbf{W}-\mathbf{\Psi}\mathbf{W}^H\mathbf{H}_{\mathrm{e}}\widehat{\mathbf{F}}-\mathbf{\Psi}\widehat{\mathbf{F}}^H\mathbf{H}_{\mathrm{e}}^H\mathbf{W}+\zeta^{-2}\sigma^2\mathbf{\Psi}\mathbf{W}^H\mathbf{W})$,  $\mathbf{W}=\mathrm{diag}({w}_{1},\cdots, {w}_{K})$, $\mathbf{\Psi} = \mathrm{diag}({\psi}_{1},\cdots, {\psi}_{K}$, and $\mathbf{H}_{\mathrm{e}} = [\mathbf{h}_{\mathrm{e},1}, \dots, \mathbf{h}_{\mathrm{e},K}]^H$. 
%where $\overline{g}=\mathrm{Tr}(\mathbf{\Psi}+\zeta^{-2}\mathbf{\Psi}{\mathbf{W}}^H\mathbf{H}_{\mathrm{e}}{\mathbf{F}}{\mathbf{F}}^H\mathbf{H}_{\mathrm{e}}^H\mathbf{W}-\zeta^{-1}\mathbf{\Psi}\mathbf{W}^H\mathbf{H}_{\mathrm{e}}{\mathbf{F}}-\zeta^{-1}\mathbf{\Psi}{\mathbf{F}}^H\mathbf{H}_{\mathrm{e}}^H\mathbf{W}+\zeta^{-2}\sigma^2\mathbf{\Psi}\mathbf{W}^H\mathbf{W})$,  $\mathbf{W}=\mathrm{diag}({w}_{1},\cdots, {w}_{K})$, $\mathbf{\Psi} = \mathrm{diag}({\psi}_{1},\cdots, {\psi}_{K}$, and $\mathbf{H}_{\mathrm{e}} = [\mathbf{h}_{\mathrm{e},1}, \dots, \mathbf{h}_{\mathrm{e},K}]^H$. 

\emph{Remark 2:} In \cite{2023_Dai_PS-DPC} and \cite{2024_Jin_PS-DPC}, the authors first considered the energy efficiency maximization problem for an IRS-assisted system with PS-DPC under a BS transmit power constraint. Different from these works, in our study, we focus on the rate maximization problem with a system power constraint including the power consumption at both the BS and the IRS, which is also of great importance in practical applications as noted in Section \ref{sec:introduction}. Furthermore, one challenge in the considered problem is that the available power needs to be divided between BS and IRS, which leads to a different design approach compared to those in \cite{2023_Dai_PS-DPC} and \cite{2024_Jin_PS-DPC}. In particular, while the AO algorithm is used in \cite{2023_Dai_PS-DPC} and \cite{2024_Jin_PS-DPC} to alternately update the beamforming at the BS and the IRS, this algorithm is not applicable in this work because it can significantly restrict the IRS's beamforming DoF, as noted in \emph{Remark 1}. %In particular, if we directly use the conventional AO method to solve this problem, the power allocation between BS and IRS cannot be effectively optimized, as only IRS configurations with fewer on-state PIN didoes than the initial states is achievable. It can significantly reduce the beamforming DoF of the IRS, leading to a quick convergence to highly sub-optimal local maximums.  %To tackle this challenge, in the subsequent subsection, we propose two novel JPABF algorithms that can jointly optimize the power allocation and the beamforming at BS and IRS.}

\subsection{JPABF Methods for Beamforming Design}\label{subsec:JPABF Algorithm for the IRS Design}
Given the optimized $\mathbf{W}$ and $\mathbf{\Psi}$, in this subsection, we focus on the optimization of $\zeta$, $\widehat{\mathbf{F}}$, and $\mathbf{b}$. Specifically, to tackle the challenging power allocation problem under the system power constraint, two novel methods are proposed to jointly optimize the power allocation and the beamforming at BS and IRS. 
\subsubsection{JPABF-$\boldsymbol{F}_{{opt}}$ method}
In this method, we first derive closed-form solutions of $\zeta$ and $\widehat{\mathbf{F}}$ for given $\mathbf{b}$. Based on the Karush-Kuhn-Tucker (KKT) conditions, the optimal $\zeta$ and $\widehat{\mathbf{F}}$ for \eqref{prb:WMMSE problem multi user1} are given by 
\begin{equation}\label{eqn:F solution}
     \widehat{\mathbf{F}} = {\mathbf{V}}^{-1}\mathbf{H}_{\mathrm{e}}^H\mathbf{W}\mathbf{\Psi}, \ \zeta=\sqrt{(P_\mathrm{0}-P_{\mathrm{PIN}}\mathbf{1}_{M}^T\mathbf{b})}\|\mathbf{\widehat{F}}\|^{-1}_F,
\end{equation}
where ${\mathbf{V}}={\mathbf{V}}^H=\big(\sigma^2/(P_\mathrm{0}-P_{\mathrm{PIN}}\mathbf{1}_{M}^T\mathbf{b})\big)\mathrm{Tr}\big(\mathbf{\Psi}\mathbf{W}^H\mathbf{W}\big)\mathbf{I}_{N}+\mathbf{H}^H_{\mathrm{e}}\mathbf{W}\mathbf{\Psi}\mathbf{W}^H\mathbf{H}_{\mathrm{e}}$. To jointly optimize the power allocation and the beamforming design at the BS and the IRS, we further substitute the optimized $\zeta$ and $\widehat{\mathbf{F}}$, and rewrite the objective function in \eqref{prb:WMMSE problem multi user1} as a function of $\mathbf{b}$. Specifically, note that
\begin{equation}\label{eqn:F property}
    \begin{array}{c}

         \mathrm{Tr}(\mathbf{\widehat{F}}^H\mathbf{V}\mathbf{\widehat{F}})=\frac{\sigma^2}{\zeta^2}\mathrm{Tr}\big(\mathbf{\Psi}\mathbf{W}^H\mathbf{W}\big)+\mathrm{Tr}(\mathbf{\widehat{F}}^H\mathbf{H}^H_{\mathrm{e}}\mathbf{W}\mathbf{\Psi}\mathbf{W}^H\mathbf{H}_{\mathrm{e}}\mathbf{\widehat{F}}),\\
          \mathbf{H}_{\mathrm{e}}^H\mathbf{W}\mathbf{\Psi}=\mathbf{V}\mathbf{\widehat{F}}.\\
    \end{array}
\end{equation}
Therefore, the objective function $\overline{g}$ in \eqref{prb:WMMSE problem multi user1} can be rewritten as
\begin{equation}\label{eqn:g reformulation}
    \begin{array}{rl}
         {\overline{g}}&= \mathrm{Tr}\big(\mathbf{\Psi}+\widehat{\mathbf{F}}^H\mathbf{H}_{\mathrm{e}}^H\mathbf{W}\mathbf{\Psi}{\mathbf{W}}^H\mathbf{H}_{\mathrm{e}}\widehat{\mathbf{F}}-\mathbf{\Psi}\mathbf{W}^H\mathbf{H}_{\mathrm{e}}\widehat{\mathbf{F}}-\\
         &\ \ \ \ \ \ \ \mathbf{\Psi}\widehat{\mathbf{F}}^H\mathbf{H}_{\mathrm{e}}^H\mathbf{W}+\frac{\sigma^2}{\zeta^2}\mathbf{\Psi}\mathbf{W}^H\mathbf{W}\big) \\
         &=\mathrm{Tr}\big(\mathbf{\Psi}+ \mathbf{\widehat{F}}^H\mathbf{V}\mathbf{\widehat{F}}-\mathbf{\widehat{F}}^H\mathbf{V}^H\mathbf{\widehat{F}}-\mathbf{\widehat{F}}^H\mathbf{V}\mathbf{\widehat{F}}\big)\\
         %& = \mathrm{Tr}\big(\mathbf{\Psi}-\mathbf{\widehat{F}}^H\mathbf{V}\mathbf{\widehat{F}}\big)\\
         & = \mathrm{Tr}\big(\mathbf{\Psi}-\mathbf{\Psi}\mathbf{W}^H\mathbf{H}_{\mathrm{e}}\mathbf{V}^{-1}\mathbf{H}_{\mathrm{e}}^H\mathbf{W}\mathbf{\Psi}\big)\\
         %& = \mathrm{Tr}\Big(\mathbf{\Psi}-\mathbf{\Psi}\mathbf{W}^H\mathbf{H}_{\mathrm{e}}\Big(\frac{\sigma^2\mathrm{Tr}\big(\mathbf{\Psi}\mathbf{W}^H\mathbf{W}\big)}{P_\mathrm{0}-P_{\mathrm{PIN}}\mathbf{1}_{M}^T\mathbf{b}}\mathbf{I}_{N}+\mathbf{H}^H_{\mathrm{e}}\mathbf{W}\mathbf{\Psi}\\
         %&\ \ \ \ \ \ \ \times\mathbf{W}^H\mathbf{H}_{\mathrm{e}}\Big)^{-1}\mathbf{H}_{\mathrm{e}}^H\mathbf{W}\mathbf{\Psi}\Big)\\
         &\overset{(\mathrm{a})}{=}\mathrm{Tr}\left(\left(\mathbf{\Psi}^{-1}+\frac{P_\mathrm{0}-P_{\mathrm{PIN}}\mathbf{1}_{M}^T\mathbf{b}}{\sigma^2\mathrm{Tr}\big(\mathbf{\Psi}\mathbf{W}^H\mathbf{W}\big)}\mathbf{W}^H\mathbf{H}_{\mathrm{e}}\mathbf{H}_{\mathrm{e}}^H\mathbf{W}\right)^{-1}\right)\overset{\wedge}{=}\widetilde{g},
    \end{array}
\end{equation}
where $(\mathrm{a})$ is derived using the Woodbury matrix identity \cite{woodbury}. Note that for the $\widetilde{g}$ specified in \eqref{eqn:g reformulation}, $\mathbf{b}$ is incorporated not only in matrix $\mathbf{H}_\mathrm{e}$ but also in coefficient $\frac{P_\mathrm{0}-P_{\mathrm{PIN}}\mathbf{1}_{M}^T\mathbf{b}}{\sigma^2\mathrm{Tr}\big(\mathbf{\Psi}\mathbf{W}^H\mathbf{W}\big)}$. Now, \eqref{prb:WMMSE problem multi user1} can be rewritten as the following problem
%and then solve the new problem with coordinate descent (CD) algorithm. Specifically,By substituting the above optimized $\zeta$ and $\widehat{\mathbf{F}}$ into the problem in \eqref{prb:WMMSE problem multi user}, the equivalent WMMSE problem reduces to the following form which is only related to $\mathbf{b}$
\begin{equation}\label{prb:WMMSE problem multi user2}
\begin{array}{cl}
\displaystyle{\minimize_{\mathbf{b}}}& \widetilde{g} \\
\mathrm{subject \; to} 
& \mathrm{C}2:\mathbf{b}\in\{0,1\}^{M},
\end{array}
\end{equation}
 where $\mathbf{b}$ is the only optimization variable. Since $\widetilde{g}$ is obtained by substituting the optimized $\zeta$ and $\widehat{\mathbf{F}}$, we can ensure that the optimized digital beamforming matrix as well as the corresponding power budget $P_{\mathrm{BS,t}}=P_\mathrm{0}-P_{\mathrm{PIN}}\mathbf{1}_{M}^T\mathbf{b}$ at the BS are updated for each possible $\mathbf{b}$. Therefore, by solving \eqref{prb:WMMSE problem multi user2}, we can facilitate the joint optimization of the system power allocation and the beamforming design at BS and IRS.

To solve \eqref{prb:WMMSE problem multi user2}, we apply the coordinate descent (CD) algorithm, and iteratively optimize each element of $\mathbf{b}$ while keeping the others fixed. Without loss of generality, we assume that the element to be optimized in the current iteration is $[\mathbf{b}]_m$, which can only be set to 1 or 0. Therefore, we respectively calculate $\widetilde{g}$ for $[\mathbf{b}]_m=1$ and $0$, i.e., $\widetilde{g}\big|_{[\mathbf{b}]_m=1}$ and $\widetilde{g}\big|_{[\mathbf{b}]_m=0}$, thus determining the optimal $[\mathbf{b}]_m$ as
\begin{equation}\label{eqn:b optimization}
 [\mathbf{b}]_m=\left\{\begin{array}{cl}
    0,&\widetilde{g}\big|_{[\mathbf{b}]_m=0}\leq\widetilde{g}\big|_{[\mathbf{b}]_m=1},  \\
         1,& \mathrm{otherwise}.
    \end{array}\right.
\end{equation}
 After the optimization of all elements of $\mathbf{b}$, we then determine the optimized $\zeta$ and $\widehat{\mathbf{F}}$ according to \eqref{eqn:F solution}. 
 
 By iteratively optimizing all variables, the overall JPABF-$\mathbf{F}_{\mathrm{opt}}$ method is summarized in \textbf{Algorithm 2}. As mentioned above, this method facilitates joint power allocation and beamforming design in IRS-assisted systems with PS-DPC. However, it requires the updating of $\widetilde{g}$, which necessitates a matrix inversion, in each iteration of the CD algorithm. Consequently, the computational complexity of this method is substantial, on the order of $\mathcal{O}(KNM^2+K^3M)$.
%However, the complexity of the JPABF-$\mathbf{F}{\mathrm{opt}}$ method is considerable due to the necessity to update $\widetilde{g}$ during each iteration of the CD algorithm.which contains the inverse of matrices, during each iteration of the CD algorithm, the complexity of the JPABF-$\mathbf{F}_{\mathrm{opt}}$ method is high, i.e., in the order of $\mathcal{O}{(K^3M^2)}$.
\begin{algorithm}[t]
\label{alg}
	\caption{JPABF-$\mathbf{F}_{\mathrm{opt}}$ Method}
	\begin{algorithmic}[1]
	\STATE Initialize $\zeta_{(0)}$, $\widehat{\mathbf{F}}_{(0)}$, $\mathbf{b}_{(0)}$.  Set $i=0$, ${g}_{(0)}=\infty$, and $\Delta$

\REPEAT	
    \STATE $i\leftarrow i+1$
    \STATE Obtain $\mathbf{W}_{(i)}$ and $\mathbf{\Psi}_{(i)}$ according to \eqref{eqn:W psi}
     \STATE Obtain ${g}_{(i)}$ according to \eqref{prb:WMMSE problem multi user}
    \STATE Obtain $\mathbf{b}_{(i)}$ by solving \eqref{prb:WMMSE problem multi user2} 
     \STATE Obtain $\zeta_{(i)}$ and $\widehat{\mathbf{F}}_{(i)}$ according to \eqref{eqn:F solution}
\UNTIL ${g}_{(i-1)}-{g}_{(i)}\leq \Delta$
\STATE Output the solutions $\mathbf{F}^\star=\zeta_{(i)}\widehat{\mathbf{F}}_{(i)}$, $\mathbf{b}^\star=\mathbf{b}_{(i)}$
	\end{algorithmic}
\end{algorithm}

\subsubsection{{JPABF-$\boldsymbol{F}_{{scale}}$ method}}
To further reduce the computational complexity, in the JPABF-$\mathbf{F}_{\mathrm{scale}}$ method, we fix $\zeta$ and $\widehat{\mathbf{F}}$ to the solutions obtained in the previous iteration (denoted by $\zeta_{\mathrm{p}}$ and $\widehat{\mathbf{F}}_{\mathrm{p}}$), and define the BS beamformer as $\mathbf{F}= \varrho\zeta_{\mathrm{p}}\widehat{\mathbf{F}}_{\mathrm{p}}$ by introducing a power scaling coefficient $\varrho\in\mathbb{R}_0^+$. Then, we only optimize $\varrho$ for each possible $\mathbf{b}$ in the CD algorithm. In this case, the beamformer at the BS, $\mathbf{F}$, can be scaled by the coefficient $\varrho$ for each possible $\mathbf{b}$, thus also achieving flexible power allocation between BS and IRS. Specifically, we reformulate  \eqref{prb:WMMSE problem multi user} into the following problem
\begin{equation}\label{prb:WMMSE problem multi user5}
\begin{array}{cl}
\displaystyle{\minimize_{{\varrho,\mathbf{b}}}} & 
\widehat{g}\\
\mathrm{subject \; to} & {\mathrm{C}1}:P_{\mathrm{PIN}}\mathbf{1}_{M}^T\mathbf{b}\leq P_0-\varrho^2\zeta_{\mathrm{p}}\mathrm{Tr}(\widehat{\mathbf{F}}_{\mathrm{p}}^H\widehat{\mathbf{F}}_{\mathrm{p}}), \\
& \mathrm{C}2:\mathbf{b}\in\{0,1\}^{M},
\end{array}
\end{equation}
with objective function
\begin{equation}\label{eqn:g}
\begin{array}{l}
\widehat{g}=\varrho^2(2\mathbf{b}-\mathbf{1}_{M})^H\mathbf{\Xi}(2\mathbf{b}-\mathbf{1}_{M})-2\varrho\mathcal{R}\big\{\boldsymbol{\rho}^H(2\mathbf{b}-\mathbf{1}_{M})\big\},
\end{array}
\end{equation}
where $\boldsymbol{\rho}={\rm{diag}}(\mathbf{H}^H\mathbf{W}\mathbf{\Psi}\widehat{\mathbf{F}}_{\mathrm{p}}^H\mathbf{G}^H)$ and $\mathbf{\Xi}=(\mathbf{H}^H\mathbf{W}\mathbf{\Psi}\mathbf{W}^H\mathbf{H})\circ(\mathbf{G}\widehat{\mathbf{F}}_{\mathrm{p}}\widehat{\mathbf{F}}_{\mathrm{p}}^H\mathbf{G}^H)^T$. The derivation of \eqref{eqn:g} eliminates the terms in \eqref{prb:WMMSE problem multi user1} that are irrelevant to the optimization variables $\varrho$ and $\mathbf{b}$, and leverages the following matrix properties: $\mathrm{Tr}(\mathbf{X+Y})=\mathrm{Tr}(\mathbf{X})+\mathrm{Tr}(\mathbf{Y})$, $\mathrm{Tr}(\mathbf{XY})=\mathrm{Tr}(\mathbf{YX})$, ${\rm{Tr}}(\mathbf{X}\mathbf{Z})=\mathbf{x}^T\mathbf{z}$, and ${\rm{Tr}}(\mathbf{Z}^H\mathbf{X}\mathbf{Z}\mathbf{Y})=\mathbf{z}^H(\mathbf{X}\circ\mathbf{Y}^T)\mathbf{z}$, which hold for any matrices $\mathbf{X}$, $\mathbf{Y}$, and diagonal matrix $\mathbf{Z}$, where $\mathbf{x}=\rm{diag}(\mathbf{X})$ and $\mathbf{z}=\rm{diag}(\mathbf{Z})$. Therefore, we can apply a similar CD algorithm to optimize each element in $\mathbf{b}$. Specifically, for a given $\mathbf{b}$, we note that \eqref{prb:WMMSE problem multi user5} is a quadratic programming problem for $\varrho$, which has a closed-form solution. Hence, during the optimization of the $m$-th element of $\mathbf{b}$, we can respectively calculate $\widehat{g}$ for $[\mathbf{b}]_m=1$ and $0$ and the corresponding optimal $\varrho$, i.e., $\widehat{g}\big|_{[\mathbf{b}]_m=1,\varrho^\star_{[\mathbf{b}]_m=1}}$ and $\widehat{g}\big|_{[\mathbf{b}]_m=0,\varrho^\star_{[\mathbf{b}]_m=0}}$, thus determining the optimal $[\mathbf{b}]_m$ similar to \eqref{eqn:b optimization}. As objective function $\widehat{g}$ is equivalent to that in the original WMMSE problem \eqref{prb:WMMSE problem multi user}, the $\mathbf{b}$ and $\mathbf{F}$ obtained by solving problem \eqref{prb:WMMSE problem multi user5} can also reduce objective function $g$ in \eqref{prb:WMMSE problem multi user}, thus ensuring the convergence of the JPABF-$\mathbf{F}_{\mathrm{scale}}$ method. After the optimization of $\mathbf{b}$, $\zeta$ and $\widehat{\mathbf{F}}$ are then optimized according to \eqref{eqn:F solution}.%Note that after each iteration, the updated $\mathbf{b}$, $\zeta$ and the fixed $\widehat{F}$ are 
\begin{algorithm}[t]
\label{alg}
	\caption{JPABF-$\mathbf{F}_{\mathrm{scale}}$ Method}
	\begin{algorithmic}[1]
	\STATE Initialize $\zeta_{(0)}$, $\widehat{\mathbf{F}}_{(0)}$, $\mathbf{b}_{(0)}$.  Set $i=0$, ${g}_{(0)}=\infty$, and $\Delta$

\REPEAT	
    \STATE $i\leftarrow i+1$
    \STATE Obtain $\mathbf{W}_{(i)}$ and $\mathbf{\Psi}_{(i)}$ according to \eqref{eqn:W psi}
     \STATE Obtain ${g}_{(i)}$ according to \eqref{prb:WMMSE problem multi user}
    \STATE Obtain $\mathbf{b}_{(i)}$ by solving \eqref{prb:WMMSE problem multi user5} %based on $\zeta_{(i-1)}$ and $\widehat{\mathbf{F}}_{(i-1)}$
     \STATE Obtain $\zeta_{(i)}$ and $\widehat{\mathbf{F}}_{(i)}$ according to \eqref{eqn:F solution}
\UNTIL ${g}_{(i-1)}-{g}_{(i)}\leq \Delta$
\STATE Output the solutions $\mathbf{F}^\star=\zeta_{(i)}\widehat{\mathbf{F}}_{(i)}$, $\mathbf{b}^\star=\mathbf{b}_{(i)}$
	\end{algorithmic}
\end{algorithm}

The JPABF-$\mathbf{F}_{\mathrm{scale}}$ method is summarized in \textbf{Algorithm 3}. Thanks to the joint optimization of power scaling variable $\varrho$ and $\mathbf{b}$, the JPABF-$\mathbf{F}_{\mathrm{scale}}$ method also facilitates flexible power division between BS and IRS during beamforming design. Furthermore, compared to the JPABF-$\mathbf{F}_{\mathrm{opt}}$ method, the JPABF-$\mathbf{F}_{\mathrm{scale}}$ method avoids the high-complexity calculations associated with matrix inversion in \eqref{eqn:g reformulation}, thus significantly reducing the computational complexity. In particular, the complexity of the JPABF-$\mathbf{F}_{\mathrm{scale}}$ method is on the order of $\mathcal{O}{(KM^2)}$, which is substantially lower than that of the JPABF-$\mathbf{F}_{\mathrm{opt}}$ method. }

\section{Simulation Results}\label{sec:Simulation Results}
\subsection{Simulation Setup}\label{subsec:Simulation Setup}

%We consider a downlink narrow-band communication system, where a BS, equipped with $N = 5$ antennas, services $3$ single-antenna user through an IRS. Specifically, both the BS and the center of the IRS are positioned $15\ \mathrm{m}$ above the user level, and the BS is located in the normal direction of the IRS. The horizontal distance between the BS and IRS is $d_{\mathrm{BS-IRS}}=20\ \mathrm{m}$, while the horizontal distance from the IRS to the $k$-th user, $d_{\mathrm{IRS-user}_k}$, follows a uniform distribution in the range of $[10\ \mathrm{m}, 120\ \mathrm{m}]$. The antenna array at the BS is assumed to be a ULA, and the IRS is assumed to be a UPA. For both arrays, the distances between neighboring units are set to $\frac{1}{2}\lambda$, with $\lambda=0.07\ \mathrm{m}$. The power consumption of a PIN diode in the on-state is $P_{\mathrm{PIN}}=12\ \mathrm{mW}$. The noise variance of the user is set to $\sigma^2=-110\ \mathrm{dBm}$.  The channel gains of $\mathbf{G}$ and $\mathbf{h}_k$ are given by $\alpha_\mathrm{G} =\sqrt{10^{-4}d_{\mathrm{BS-IRS}}^{-2.2}}$ and $\alpha_\mathrm{h} = \sqrt{10^{-4}d_{\mathrm{IRS-user}_k}^{-2.2}} $, respectively \cite{2022_Mu_STAR-RIS}. The Rician factor $\kappa$ is set to 10.%All simulation results are obtained by averaging over 500 independent realizations.
We consider a downlink narrowband IRS-assisted communication system, where the BS is equipped with $N = 5$ antennas. The BS is located in the normal direction of the IRS. The distance between BS and IRS is $d_{\mathrm{BS-IRS}}=20\ \mathrm{m}$, while the distance from the IRS to the $k$-th user, $d_{\mathrm{IRS-UE}_k}$, follows a uniform distribution in the range of $[50\ \mathrm{m}, 70\ \mathrm{m}]$. %The power consumption of a PIN diode in the on-state is $P_{\mathrm{PIN}}=12\ \mathrm{mW}$ \cite{2022_Arxiv_Cui_dynamic_power_consumption}. 
The other system parameters are presented in Table I.
\begin{table}[H]\centering
\caption{Simulation parameters}
\begin{tabular}{|l|l|l|}
\hline$P_{\mathrm{PIN}}$ & Power consumption of an on-state PIN diode & $12\ \mathrm{mW}$ \cite{2022_Arxiv_Cui_dynamic_power_consumption} \\
\hline$\lambda$ & Carrier wavelength & $0.07\ \mathrm{m}$ \\
\hline$\sigma^2$ & Noise power at users & $-110\ \mathrm{dBm}$ \\
\hline$\mathrm{PL}_\mathrm{G}$ & Path loss of the BS-IRS channel & ${10^{-4}d_{\mathrm{BS-IRS}}^{-2.2}}$ \\
\hline$\mathrm{PL}_{\mathrm{h}_k}$ & Path loss of the IRS-UE$_k$ channel & ${10^{-4}d_{\mathrm{IRS-UE}_k}^{-2.2}}$ \\
\hline$\vartheta_{\mathrm{t},k}$ & the elevation AoD of $\mathbf{h}_{\mathrm{LoS}}$ & $[0,\frac{\pi}{4}]$ \\
\hline$\psi_{\mathrm{t},k}$ & the azimuth AoD of $\mathbf{h}_{\mathrm{LoS}}$ & $[0,2\pi)$ \\
\hline$\kappa_{\mathrm{G}}$ & Rician factors of the BS-IRS channel& $8$ \\
\hline$\kappa_{\mathrm{h}_k}$ & Rician factors of the IRS-UE$_k$ channel& $8$ \\
\hline$\Delta$ & the convergence tolerance& $0.005$ \\
%\hline$\kappa$ & Rician factor  & $10\ \mathrm{dB}$ \\
\hline
\end{tabular}
\end{table}
%\begin{equation}
 %   \begin{array}{cc}
  %      \alpha_\mathrm{G} =\sqrt{10^{-3}d_{\mathrm{BS-IRS}}^{-2.2}},   \\
  %       \alpha_\mathrm{h} = \sqrt{10^{-3}d_{\mathrm{IRS-UE}_k}^{-2.2}} .
  %  \end{array}
%\end{equation}

%For the multi-user case, the system is configured to support 3 UEs, i.e., $K=3$. Additionally, the BS-IRS channel $\mathbf{G}$ is assumed to be a Rician channel with a Rician factor of $4$. The other simulation parameters remain consistent with those outlined for the single-user case.
%All simulation results are obtained by averaging over 200 independent realizations.

\subsection{Baseline Methods}
To illustrate the efficiency of the proposed methods, we adopt the following baseline methods for comparison:
\begin{itemize}
    \item \textbf{AO, random initialization \big(AO (rand init)\big)}: This method utilizes the conventional AO approach to iteratively optimize the BS beamformer and the IRS configuration. In each iteration, one variable is optimized while the other variables are kept fixed. For the initial state of $\mathbf{b}$, $\mathbf{b}_0$, we randomly select the number and locations of the on-state PIN diodes such that the system power constraint is satisfied, i.e., $P_{\mathrm{PIN}}\mathbf{1}_{M}^T\mathbf{b}_{(0)}\leq P_0$. Therefore, when $P_0 \leq P_{\mathrm{PIN}}M$, the power allocation between BS and IRS is approximately equal in the initial state. If $P_0 > P_{\mathrm{PIN}}M$, the initial state $\mathbf{b}_{(0)}$ is chosen without constraints, leading to approximately half of the PIN diodes being turned on.%Utilizing the conventional AO approach, this method iteratively optimizes the BS beamformer and IRS configuration for the single-user case as defined in problem \eqref{eqn:problem formulation} and for the multi-user case as in problem \eqref{prb:WMMSE problem multi user}. In each iteration, one variable is optimized while the other variables are kept fixed. For the initial state of $\mathbf{b}$, $\mathbf{b}_0$, we randomly select the number and locations of the on-state PIN diodes such that the system power constraint is satisfied, i.e., $P_{\mathrm{PIN}}\mathbf{1}_{M}^T\mathbf{b}_{(0)}\leq P_0$. Therefore, when $P_0 \leq P_{\mathrm{PIN}}M$, the power allocation between the BS and IRS is approximately equal in the initial state. If $P_0 > P_{\mathrm{PIN}}M$, the initial state $\mathbf{b}_{(0)}$ is chosen without constraints, leading to approximately half of the PIN diodes being turned on.
    \item \textbf{AO, zero initialization \big(AO (zero init)\big)}: This method also applies the AO algorithm for beamforming optimization at BS and IRS. In contrast to the AO (rand init) method, all PIN diodes are initially set to the off-state.
    %\item Random phase shift (rand PS): In this method, the elements in $\mathbf{b}$ are randomly set to 1 or 0 without any optimization. The BS beamforming is then optimized to maximize the rate for the given random $\mathbf{b}$.
    \item \textbf{Ignore PS-DPC}: In this method, we optimize the system beamforming with the proposed methods while disregarding the PS-DPC at the IRS. Therefore, this method aims to enhance the IRS beamforming quality as much as possible without considering the PS-DPC. For the multi-user case, we respectively use the JPABF-$\mathbf{F}_{\mathrm{opt}}$ and JPABF-$\mathbf{F}_{\mathrm{scale}}$ methods to optimize the beamforming designs at the BS and the IRS while disregarding the PS-DPC, which are referred to as Ignore PS-DPC ($\mathbf{F}_{\mathrm{opt}}$) and Ignore PS-DPC ($\mathbf{F}_{\mathrm{scale}}$), respectively.
    \item \textbf{Semidefinite programming (SDP) relaxation} \cite{2023_Dai_PS-DPC}: This method aims to enhance the energy efficiency in multi-user IRS-assisted systems with PS-DPC for a power constraint at the BS, and utilizes the AO algorithm to optimize the beamforming at the BS and the IRS alternately. At the BS, beamforming is designed using the zero-forcing technique, while at the IRS, it is optimized via an SDP relaxation algorithm. %It is critical to highlight that, in fact, the SDP relaxation method is unable to address the total system power constraint, which is a key focus of this paper. Consequently, is constrained to demonstrating its beamforming performance under the BS's power constraints, facilitating a comparative analysis with the methods proposed in this paper. Specifically, the SDP relaxation algorithm is initially employed to design beamforming under the BS's power constraint. Then, we calculate its total system power $P_0$ by summing the consumptions at the BS and the IRS.}

    %As noted in \emph{Remark 2}, this method cannot address the total system power constraint considered in this paper in fact, so here, we just show the simulation results of this method under the BS's power constraint in order to compare the beamforming performance of this method with the proposed methods. In particular, for the SDP relaxation algorithm, we initially use this method to design beamforming under the BS's power constraint, and then calculate total system power $P_0$ by summing the consumptions at the BS and the IRS.}
\end{itemize}

\subsection{Single-User: System Performance versus Power Budget}\label{subsec:Single-UE: System Performance with Different System Power}
\begin{figure}
	\centering
	\includegraphics[width=0.28\textwidth]{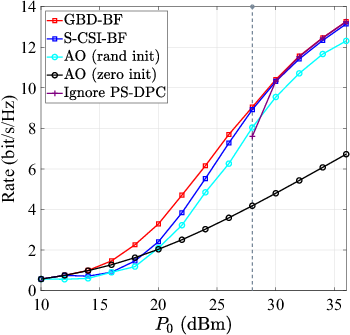}
	\caption{Rate versus $P_{\mathrm{0}}$ for different beamforming methods.}\label{fig:Simulation_1_SU_R-P}
\end{figure}
We first evaluate the performance of the proposed methods in the single-user case. In Fig. \ref{fig:Simulation_1_SU_R-P}, we plot the downlink rate versus the system power budget $P_0$ for $M=100$. When $P_0< P_{\mathrm{PIN}}= 10.8\ \mathrm{dBm}$, the system power is insufficient to turn on any PIN diodes, leading to a scenario where $\mathbf{b}=\mathbf{0}_M$ and only the BS beamforming can be optimized. Consequently, in this low-power regime, all considered beamforming methods converge to a similar rate performance. When $P_0> 10.8\ \mathrm{dBm}$, the proposed GBD-BF method consistently achieves the highest rate. This superior performance is attributed to the method's ability to jointly optimize power allocation and beamforming at the BS and the IRS while taking into account the PS-DPC. On the other hand, the proposed S-CSI-BF method yields competitive performance at high values of $P_0$, but its effectiveness diminishes at low values of $P_0$. This is primarily because, when $P_0$ is small, the $t^\star$ obtained in \eqref{eqn:zero point of SNR derivative} is close to $\frac{\pi}{2}$. Consequently, $\mathbb{E}\left\{\mathcal{R}\left\{{h}_{\mathrm{o}}{\varphi}^\star\right\}\right\} = \frac{2}{M\pi}\cos{t}$ is also close to zero. In this scenario, the assumption used in the S-CSI-BF method that $\mathbb{E}\left\{\mathcal{I}\left\{{h}_{\mathrm{o}}{\varphi}^\star\right\}\right\}$ is negligible compared to $\mathbb{E}\left\{\mathcal{R}\left\{{h}_{\mathrm{o}}{\varphi}^\star\right\}\right\}$, is no longer fully satisfied, which leads to some performance loss. Nonetheless, the S-CSI-BF method remains highly attractive due to its much lower complexity compared to the GBD-BF method, making it an efficient choice in practical applications. %{\color{black}On the other hand, the proposed S-CSI-BF method yields competitive performance at high values of $P_0$, but its effectiveness diminishes at low values of $P_0$. This degradation in performance is mainly because, when the power budget is limited, the $t^\star$ obtained in \eqref{eqn:zero point of SNR derivative} is closed to $\frac{\pi}{2}$, which leads to a $\mathbb{E}\left\{\mathcal{R}\left\{{h}_{\mathrm{o}}{\varphi}^\star\right\}\right\}=\frac{2}{M\pi}\cos{t}$ closed to zero. In this case, the assumption $\mathbb{E}\left\{\mathcal{R}\left\{{h}_{\mathrm{o}}{\varphi}^\star\right\}\right\}>>\mathbb{E}\left\{\mathcal{I}\left\{{h}_{\mathrm{o}=0}{\varphi}^\star\right\}\right\}$ is not perfectly satisfied, which cause a notable performance loss.}%This degradation in performance is due to the rounding of the phase shifts in \eqref{eqn:BF of CSC-BF} and the approximations in \eqref{eqn:channel model approximate} and \eqref{eqn:received SNR with channel model3}, which cause a loss especially when the power budget is limited.

The conventional AO methods are susceptible to be trapped in local optima close to the initial points, as we have discussed in \emph{Remark 1}. Specifically, for the AO (zero init) method, $\mathbf{b}$ remains stuck at $\mathbf{b}_{(0)}=\mathbf{0}_M$. This is beneficial when $P_0$ is low, because turning on PIN diodes would consume a significant portion of the power without substantially improving the IRS's beamforming capability. However, as $P_0$ increases, the AO (zero init) method experiences a considerable decrease in performance since it cannot leverage the IRS's beamforming to improve the system rate performance. The AO (rand init) method, conversely, tends to divide the available power evenly between BS and IRS when $P_0\leq28\ \mathrm{dBm}$. As a result, even at low $P_0$, some PIN diodes remain in the on-state, which can lead to a significant reduction in the transmitted signal power and thus cause a noticeable performance drop.

Finally, as the Ignore PS-DPC method disregards the PS-DPC, it always turns on approximately half of the PIN diodes, leading to a PS-DPC equal to $27.8\ \mathrm{dBm}$. As a result, this method is not feasible when $P_0< 28\ \mathrm{dBm}$. For $P_0\geq 28\ \mathrm{dBm}$, the Ignore PS-DPC method exhibits lower rates than the proposed methods when the PS-DPC is the dominant factor in the system power budget, specifically for $P_0=28\ \mathrm{dBm}$. This is because it disregards the PS-DPC during the optimization of the IRS. However, as $P_0$ continues to increase, the BS becomes the primary consumer of system power. In this high-power regime, the performance of the Ignore PS-DPC method becomes comparable to that of the proposed methods, as the impact of ignoring the PS-DPC diminishes.

\begin{figure}
	\centering
 \includegraphics[width=0.31\textwidth]{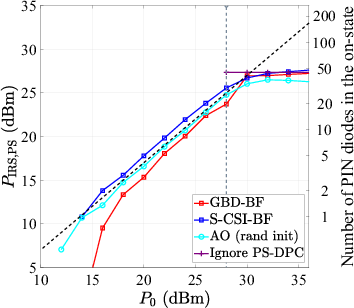}
	\caption{$P_{\mathrm{IRS,PS}}$ versus $P_{\mathrm{0}}$ for different beamforming methods.}\label{fig:Simulation_2_SU_R-M}% The dashed lines represent $P_{\mathrm{BS,t}}$, while the solid lines represent $P_{\mathrm{IRS,PS}}$.
\end{figure}
To further investigate the power allocation strategies of different methods, we show PS-DPC at the IRS ($P_{\mathrm{IRS,PS}}$) for different system power budgets in Fig. \ref{fig:Simulation_2_SU_R-M}\footnote{Since $P_{\mathrm{IRS,PS}}$ is always 0 for the AO (zero init) method, it is not shown in Fig. \ref{fig:Simulation_2_SU_R-M}.}. The transmit power of the BS, $P_{\mathrm{BS,t}}$, can be obtained from $P_{\mathrm{BS,t}}=P_{0}-P_{\mathrm{IRS,PS}}$. The black dashed line represents the scenario where the power is equally divided between BS and IRS, i.e., $P_{\mathrm{BS,t}}=P_{\mathrm{IRS,PS}}=\frac{1}{2}P_{0}$. If a method prefers to allocate more power to the IRS, its $P_{\mathrm{IRS,PS}}$ will exceed the black dashed line, and conversely for less allocated power. As can be observed, conventional AO methods cannot flexibly divide the available power between BS and IRS. In particular, the AO (rand init) method always tends to split the power evenly between BS and IRS when $P_0\leq28\ \mathrm{dBm}$, as its solutions are always trapped at local optima close to the initial states. Similarly, the AO (zero init) method consistently turns off all IRS PIN diodes for all values of $P_0$, hence allocating all of the system power to the BS. In comparison to the conventional AO methods, the proposed methods demonstrate a more flexible power allocation strategy. Specifically, for lower system power budgets, e.g., $P_0\leq28\ \mathrm{dBm}$, the proposed GBD-BF method favors allocating more power to the BS to enhance the strength of the transmitted signal, reflected in a $P_{\mathrm{IRS,PS}}$ below the black dashed line. {\color{black}However, when $P_0\ge 32\ \mathrm{dBm}$, the system power budget is higher than what is needed to meet the PS-DPC of the IRS. In this case, the PS-DPC is no longer a dominant factor in the system power budget, and the proposed methods can maximize the beamforming quality of the IRS by turning on about half of the PIN diodes. This results in a constant $P_{\mathrm{IRS,PS}}$ when $P_0\ge 32\ \mathrm{dBm}$.} Finally, as the Ignore PS-DPC method does not take into account the PS-DPC, it always turns on about half of the PIN diodes, leading to a constant PS-DPC regardless of the system power budget.

\subsection{Single-User: System Performance versus the Size of IRS}\label{subsec:Single-UE: System Performance versus the Size of IRS}
\begin{figure}
	\centering
	\includegraphics[width=0.28\textwidth]{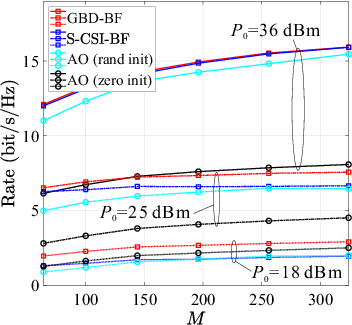}
	\caption{Rate versus $M$ for different beamforming methods.}\label{fig:Simulation_3_SU_R_M}
\end{figure}
%\footnote{As the Ignore PS-DPC method is infeasible for all $M$ considered in Fig. \ref{fig:Simulation_3_SU_R_M} when the $P_0$ is low, we do not show its performance .}
In Fig. \ref{fig:Simulation_3_SU_R_M}, we further show the relationship between the achievable rate and the number of IRS elements. As can be observed, the performance of all considered methods monotonically improves as the system power budget and the IRS size increase. Specifically, the proposed GBD-BF method achieves the highest rate for all considered values of $M$ and $P_0$ due to its capability of efficiently balancing the power consumption at BS and IRS for beamforming optimization. The proposed S-CSI-BF method approaches the performance of the GBD-BF method when the system power budget is high, but suffers from a performance loss for small values of $P_0$ and large values of $M$. Similar to Fig. \ref{fig:Simulation_1_SU_R-P}, this is because the assumption that $\mathbb{E}\left\{\mathcal{I}\left\{{h}_{\mathrm{o}}{\varphi}^\star\right\}\right\}$ is negligible compared to $\mathbb{E}\left\{\mathcal{R}\left\{{h}_{\mathrm{o}}{\varphi}^\star\right\}\right\}$, is no longer fully satisfied in these cases. For the AO (rand init) and AO (zero init) methods, as we mentioned above, their solutions are always trapped in local optimal points around the initial states. Therefore, the AO (rand init) method performs better for high $P_0$, as it can turn on some of the PIN diodes for better beamforming quality. However, when $P_0$ is very low, turning on PIN diodes cannot significantly improve the beamforming quality of the IRS, but will severely reduce the transmit power at the BS. Therefore, the AO (zero init) method outperforms the AO (rand init) method when $P_0=18\ \mathrm{dBm}$.

%This is first because a larger IRS can reflect more power to the UE, resulting in an increased power gain in the UE received signal. Furthermore, the larger number of scattering elements can also improve the beamforming quality of the IRS, and thus increase the SNR at the UE. Specifically, the proposed GBD-BF method achieves the highest rate for all $M$ and $P_0$, as it can efficiently balance the power consumption of the BS and IRS. Similar to that in Fig. \ref{fig:Simulation_1_SU_R-P}, the CSC-BF method can achieve comparable performance as that of the GBD-BF method when $P_0$ is high, but suffers a performance loss for small $P_0$. For the AO random-initial and the AO zero initializations, as we mentioned above, their solutions are always trapped at local optimal points around the initial states. Therefore, the AO random-initial performs better for high $P_0$, as it can turn on some of the PIN diodes for better beamforming quality. However, when $P_0$ is very low, turning on PIN diodes cannot significantly improve the beamforming quality of the IRS, but will severely reduce the strength of the transmitted signal at the BS. Therefore, the AO zero initialization outperforms the AO random initialization for $P_0=18\ \mathrm{dBm}$.

\begin{figure}
	\centering
	\includegraphics[width=0.31\textwidth]{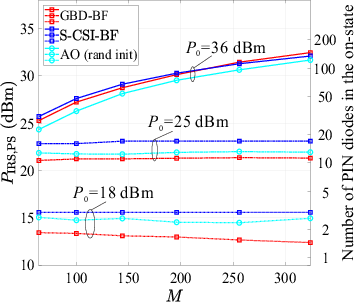}
	\caption{$P_{\mathrm{IRS,PS}}$ versus $M$ for different beamforming methods.}\label{fig:Simulation_4_SU_EI_M}
\end{figure}
%\footnote{Note that the AO (zero init) tends to turn off all PIN diodes for all cases, i.e., $P_{\mathrm{IRS,PS}}=0$, hence it is omitted in Fig. \ref{fig:Simulation_4_SU_EI_M}.}
Next, in Fig. \ref{fig:Simulation_4_SU_EI_M}, we show $P_{\mathrm{IRS,PS}}$ as a function of $M$.  As can be observed, the PS-DPC of all methods depends on the system power budget. When the system power budget is substantial, e.g., $P_0=36\ \mathrm{dBm}$, the PS-DPC is not the dominant factor in the system power consumption. In this case, the IRS can improve the beamforming quality by turning on more PIN diodes without significantly affecting the BS transmit power. As a result, the PS-DPC monotonically increases with $M$ in this case. Conversely, when the system power is limited, turning on more PIN diodes at the IRS can severely affect the power of the transmitted signals. Therefore, for $P_0=25\ \mathrm{dBm}$ and $P_0=18\ \mathrm{dBm}$, $P_{\mathrm{IRS,PS}}$ remains almost constant regardless of the IRS size. Specifically, for the proposed GBD-BF method, the PS-DPC may even slightly decrease for larger IRSs when $P_0=18\ \mathrm{dBm}$. This can be explained as follows: First of all, when the total power available is very limited, the GBD-BF method tends to allocate more power to the BS as observed in Fig. \ref{fig:Simulation_2_SU_R-M}. Given the low power allocated to the IRS, it can only support a small number of PIN diodes in the on-state. In this case, for a large IRS, the impact of altering the states of a few elements is relatively minor due to the extensive number of elements involved. Consequently, turning on only a small number of PIN diodes does not substantially increase the beamforming gain. In contrast, for a smaller IRS, turning on the same number of PIN diodes can lead to a more pronounced improvement in beamforming effectiveness, given the same PS-DPC budget. Therefore, when the total system power consumption is limited, the proposed GBD-BF method allocates more power to an IRS with a smaller size compared to that with a larger size.
%{\color{black}It is because, for a large IRS, turning on only a few PIN diodes does not significantly enhance the beamforming gain due to the vast number of elements involved. In contrast, for a smaller IRS, turning on the same number of PIN diodes results in a relatively higher improvement in beamforming effectiveness, given the same PS-DPC budget. Consequently, when the total system power consumption is fixed, a larger IRS tends to allocate more power to the BS to amplify the transmitted signal, which in turn results in a lower power consumption at the IRS, denoted as $P_{\mathrm{IRS,PS}}$.}
%an IRS with more elements, the effect of changing the phase shift configuration of an element in enhancing the channel gain is relatively weaker than that in enhancing in a smaller IRS channel gain the This indicates that for an IRS with more elements, the GBD-BF method even prefers to turn off more PIN diodes, thereby allocating more power to the BS to strengthen the transmitted signal. It is because, 
\subsection{Multi-User: System Performance versus Power Budget}\label{Multi-UE: System Performance versus Power Budget}
\begin{figure}
	\centering
	\includegraphics[width=0.28\textwidth]{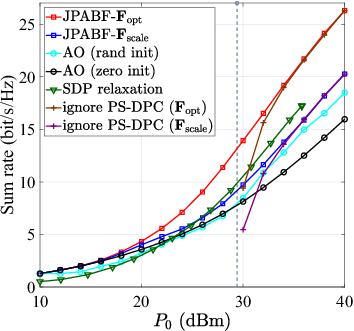}
	\caption{Sum rate versus $P_{\mathrm{0}}$ for different beamforming methods.}\label{fig:Simulation_6_MU_R-P}
\end{figure}
Now, we evaluate the performance of the proposed methods in the multi-user case. Fig. \ref{fig:Simulation_6_MU_R-P} illustrates the sum rate as a function of $P_0$ when $M=144$ and $K=3$. As can be observed, the proposed JPABF-$\mathbf{F}_{\mathrm{opt}}$ method consistently outperforms the baseline methods for all considered $P_0$, thanks to its capability of jointly optimizing the power allocation and the beamforming at the BS and the IRS. While the JPABF-$\mathbf{F}_{\mathrm{scale}}$ method exhibits lower performance than the JPABF-$\mathbf{F}_{\mathrm{opt}}$ method, its complexity is also notably lower, as discussed in Section \ref{sec:PF-WMMSE}. 
%{\color{black}Regarding the SDP relaxation method in \cite{2023_Dai_PS-DPC}, it can only operate under the power constraint at the BS, unlike the system power constraint applied in other methods. To ensure consistency in our comparisons, we align the BS power budget for the SDP relaxation method with the system power budget designated for other methods at each simulation point. Consequently, by further adding the PS-DPC at the IRS, the system power $P_0$ of the SDP relaxation method is consistently higher than that of the other methods at each simulation point. Meanwhile, its performance in terms of rate does not further improve for $P_0 > 36\ \mathrm{dBm}$. This plateau occurs because the SDP relaxation method focuses on maximizing energy efficiency, which peaks around $P_0 = 36\ \mathrm{dBm}$. Additionally, it consistently underperforms compared to the JPABF-$\mathbf{F}{\mathrm{opt}}$ method, mainly because it employs a less effective zero-forcing algorithm compared to the proposed WMMSE approach. Finally, the complexity of the SDP relaxation method is extremely high ($\mathcal{O}(M^9)$), which is significantly greater than those of the JPABF-$\mathbf{F}_{\mathrm{opt}}$ ($\mathcal{O}(K^3M^2)$) and JPABF-$\mathbf{F}_{\mathrm{scale}}$ ($\mathcal{O}(M^2)$) methods.}

{Regarding the SDP relaxation method from \cite{2023_Dai_PS-DPC}, it operates under a BS transmit power constraint rather than a system power constraint. To ensure a fair comparison, we first utilize the SDP relaxation method to design the beamforming under the BS transmit power constraint. Subsequently, we calculate the total system power $P_0$ by summing the powers consumed at both the BS and the IRS. {\color{black}Notably, the SDP relaxation method focuses on maximizing energy efficiency, which peaks around $P_0 = 36\ \mathrm{dBm}$. As a result, even with a larger power consumption budget, it maintains a constant system power consumption around $P_0 = 36\ \mathrm{dBm}$. Therefore, this method fails completely for $P_0 > 36\ \mathrm{dBm}$.} {\color{black}Furthermore, it consistently performs worse than the JPABF-$\mathbf{F}_{\mathrm{opt}}$ method, mainly because the SDP relaxation method uses a less effective zero-forcing algorithm compared to the WMMSE approach employed in this paper.} It is also worth noting that the complexity of the SDP relaxation method is very high ($\mathcal{O}(M^9)$), and is notably larger than those of the JPABF-$\mathbf{F}_{\mathrm{opt}}$ ($\mathcal{O}(KNM^2+K^3M)$) and JPABF-$\mathbf{F}_{\mathrm{scale}}$ ($\mathcal{O}(KM^2)$) methods.}
%{\color{black}Regarding the SDP relaxation method in \cite{2023_Dai_PS-DPC}, it does not consider the system power constraint but only the BS transmit power constraint. To ensure the consistency in the comparison, we align the BS power budget for the SDP relaxation method with the system power budget designated for other methods at each simulation point. Meanwhile, its performance in terms of rate does not further improve for $P_0 > 36\ \mathrm{dBm}$. This plateau occurs because the SDP relaxation method focuses on maximizing energy efficiency, which peaks around $P_0 = 36\ \mathrm{dBm}$. Additionally, it consistently underperforms the JPABF-$\mathbf{F}{\mathrm{opt}}$ method, mainly because it employs a less effective zero-forcing precoding algorithm at the BS. Finally, the complexity of the SDP relaxation method is very high ($\mathcal{O}(M^9)$), which is significantly greater than those of the JPABF-$\mathbf{F}_{\mathrm{opt}}$ ($\mathcal{O}(K^3M^2)$) and JPABF-$\mathbf{F}_{\mathrm{scale}}$ ($\mathcal{O}(M^2)$) methods.}

Similar to the single-user case, the conventional AO methods tend to get trapped in suboptimal solutions near their initial values. Therefore, they have significant performance losses compared to the proposed JPABF methods, especially when $P_0$ is high for the AO (zero init) method, and when $P_0$ is low for the AO (rand init) method. 

Finally, the Ignore PS-DPC methods show a similar trend as in the single-user case considered in Fig. \ref{fig:Simulation_1_SU_R-P}: These methods are not viable when $P_0<29.4\ \mathrm{dBm}$, as about half of the PIN diodes are turned on. They can achieve comparable performance as the proposed methods for large $P_0$ values but suffer a clear performance drop when PS-DPC dominates the system power budget, i.e., for $P_0=30\ \mathrm{dBm}$.
\subsection{Multi-User: System Performance versus the Size of IRS}\label{Multi-UE: System Performance versus the Size of IRS}
\begin{figure}
	\centering
	\includegraphics[width=0.28\textwidth]{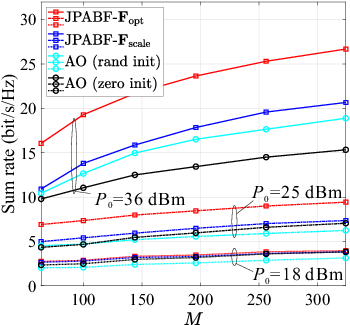}
	\caption{Sum rate versus $M$ for different beamforming methods.}\label{fig:Simulation_7_MU_R_M}
\end{figure}

%\footnote{As the complexity of the SDP relaxation method is prohibitive for IRSs with large sizes, we do not show its performance in Fig. \ref{fig:Simulation_7_MU_R_M}}
In Fig. \ref{fig:Simulation_7_MU_R_M}, we show the sum rate versus the number of IRS elements. It can be observed that the proposed JPABF methods outperform the baseline methods for all considered values of $M$. Similar to Fig. \ref{fig:Simulation_3_SU_R_M}, the performance gap between the JPABF methods and the AO (zero init) method diminishes for low $P_0$. In this situation, the JPABF methods also turn off most of the PIN diodes due to the small amount of power available.
\begin{figure}
	\centering
	\includegraphics[width=0.31\textwidth]{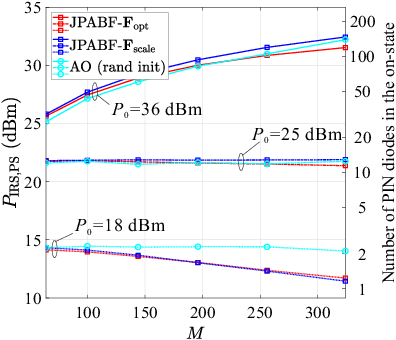}
	\caption{$P_{\mathrm{IRS,PS}}$  versus $M$ for different beamforming methods.}\label{fig:Simulation_8_MU_EI_M}
\end{figure}

In Fig. \ref{fig:Simulation_8_MU_EI_M}, we further study the PS-DPC for different IRS sizes. As can be observed, similar power allocation strategies as for the single-user case in Fig. \ref{fig:Simulation_4_SU_EI_M} can also be observed for the JPABF methods in the multi-user case. When the system power budget is sufficient, i.e., $P_0=36\ \mathrm{dBm}$, the IRS can efficiently design its phase shifts. Therefore, $P_{\mathrm{IRS,PS}}$ monotonically increases with the IRS size. However, in scenarios where the system power budget is more limited, e.g., when $P_0=18\ \mathrm{dBm}$, turning on more PIN diodes can notably reduce the power of the transmitted signals. In this case, an IRS with a larger size even sets more PIN diodes to the off-state, similar to the single-user case in Fig. \ref{fig:Simulation_4_SU_EI_M}.

\section{Conclusion}\label{sec:conclusion}
In this paper, we investigated the rate maximization optimization problem for IRS-assisted systems with PS-DPC. Considering a total system power constraint, the power allocation has been jointly optimized with the beamforming design at the BS and the IRS, leading to a favorable balance between the BS transmit power and the PS-DPC at the IRS. Specifically, for the single-user case, we proposed a GBD-BF method to jointly optimize the power allocation and the beamforming at both BS and IRS. To further reduce the computational complexity, we proposed an S-CSI-BF method, where the optimized power allocation strategy is obtained offline based on S-CSI. For the multi-user case, two JPABF methods were proposed, which could exploit the DoFs introduced by joint BS-IRS power allocation during BS and IRS beamforming optimization based on the WMMSE approach. Simulation results revealed that the proposed methods can flexibly optimize the power allocated to BS and IRS, and thus achieve higher rates than several baseline methods. The obtained power allocation strategies depend on the available system power budget. When the power budget is high, the PS-DPC is not the limiting factor, and the  IRS can turn on as many PIN diodes as necessary to achieve optimal beamforming quality. Conversely, for a limited power budget, more power is allocated to the BS to strengthen the transmitted signal, resulting in a reduced PS-DPC budget and, consequently, a lower IRS beamforming quality. %Specifically, if the system power budget is low, our simulation results indicated that the IRS with more elements can even turn to sacrifice the capability of providing higher beamforming gain, setting more PIN diodes to the off-state, since this significantly reduces the PS-DPC.


\begin{thebibliography}{00}
\bibitem{2023_Globecom_Wu_RIS_PSD-PC}
Q. Wu, T. Lin, and Y. Zhu, ``Green beamforming design for IRS-aided systems under phase shift-related power consumption,'' in \emph{Proc. IEEE Global Commun. Conf. (GLOBECOM),} Dec. 2023, pp. 613-618.

\bibitem{2020_TCCN_Li_RIS_review}
M. A. ElMossallamy, H. Zhang, L. Song, K. G. Seddik, Z. Han, and G. Y. Li, ``Reconfigurable intelligent surfaces for wireless communications: Principles, challenges, and opportunities,'' \emph{IEEE Trans. Cogn. Commun. Netw.,} vol. 6, no. 3, pp. 990-1002, Sept. 2020.

\bibitem{2021_TC_Zhang_IRS_tutorial}
Q. Wu, S. Zhang, B. Zheng, C. You, and R. Zhang, ``Intelligent reflecting surface-aided wireless communications: A tutorial,'' \emph{IEEE Trans. Commun.,} vol. 69, no. 5, pp. 3313-3351, May 2021.

\bibitem{2020_JSAC_Renzo_Smart_radio_environments_review}
M. Di Renzo et al., ``Smart radio environments empowered by reconfigurable intelligent surfaces: How it works, state of research, and the road ahead,'' \emph{IEEE J. Sel. Areas Commun.,} vol. 38, no. 11, pp. 2450-2525, Nov. 2020.

\bibitem{2019_TWC_IRS_MIMO_1}
Q. Wu and R. Zhang, ``Intelligent reflecting surface enhanced wireless network via joint active and passive beamforming,'' \emph{IEEE Trans. Wireless Commun.,} vol. 18, no. 11, pp. 5394-5409, Nov. 2019.

\bibitem{2021_TWC_IRS_MIMO_2}
P. Wang, J. Fang, L. Dai, and H. Li, ``Joint transceiver and large intelligent surface design for massive MIMO mmWave systems,'' \emph{IEEE Trans. Wireless Commun.,} vol. 20, no. 2, pp. 1052-1064, Feb. 2021.

\bibitem{2021_WC_IRS_UAV_1}
X. Pang, M. Sheng, N. Zhao, J. Tang, D. Niyato, and K. -K. Wong, ``When UAV meets IRS: Expanding air-ground networks via passive reflection,'' \emph{IEEE Wireless Commun.,} vol. 28, no. 5, pp. 164-170, Oct. 2021.

\bibitem{2021_TWC_IRS_UAV_2}
Z. Wei et al., ``Sum-rate maximization for IRS-assisted UAV OFDMA communication systems,'' \emph{IEEE Trans. Wireless Commun.,} vol. 20, no. 4, pp. 2530-2550, Apr. 2021.

\bibitem{2021_JSAC_IRS_security_1}
X. Yu, D. Xu, Y. Sun, D. W. K. Ng, and R. Schober ``Robust and secure wireless communications via intelligent reflecting surfaces,'' \emph{IEEE J. Sel. Areas Commun.,} vol. 38, no. 11, pp. 2637-2652, Nov. 2020.

\bibitem{2019_WCL_IRS_security_2}
M. Cui, G. Zhang, and R. Zhang ``Secure wireless communication via intelligent reflecting surface,'' \emph{IEEE Wireless Commun. Lett.,} vol. 8, no. 5, pp. 1410-1414, Oct. 2019.

%\bibitem{2021_TC_Yu_green_com}
%X. Yu, D. Xu, D. W. K. Ng, and R. Schober, ``IRS-assisted green communication systems: Provable convergence and robust optimization,'' \emph{IEEE Trans. Commun.,}, vol. 69, no. 9, pp. 6313-6329, Sept. 2021.


%\bibitem{2021_TWC_Dai_IRS_beamforming_data_rate}
%P. Wang, J. Fang, L. Dai, and H. Li, "Joint transceiver and large intelligent surface design for massive MIMO mmWave systems," \emph{IEEE Trans. Wireless Commun.,} vol. 20, no. 2, pp. 1052-1064, Feb. 2021.

%\bibitem{2020_WCL_Renzo_SE_EE_with_hardware_impairments}
%S. Zhou, W. Xu, K. Wang, M. Di Renzo, and M. -S. Alouini, "Spectral and energy efficiency of IRS-assisted MISO communication with hardware impairments," \emph{IEEE Wireless Commun. Lett.,} vol. 9, no. 9, pp. 1366-1369, Sept. 2020.

%\bibitem{2021_Tcom_Larsson_SE}
%J. Yuan, Y. -C. Liang, J. Joung, G. Feng, and E. G. Larsson, "Intelligent reflecting surface-assisted cognitive radio system," \emph{IEEE Trans. Commun.,} vol. 69, no. 1, pp. 675-687, Jan. 2021.

\bibitem{2019_TWC_Yuen_RIS_EE}
C. Huang, A. Zappone, G. C. Alexandropoulos, M. Debbah, and C. Yuen, ``Reconfigurable intelligent surfaces for energy efficiency in wireless communication,'' \emph{IEEE Trans. Wireless Commun.,} vol. 18, no. 8, pp. 4157-4170, Aug. 2019.

\bibitem{2022_TWC_Cui_EE_disctributeD_IRS}
Z. Yang et al, ``Energy-efficient wireless communications with distributed reconfigurable intelligent surfaces,'' \emph{IEEE Trans. Wireless Commun.,} vol. 21, no. 1, pp. 665-679, Jan. 2022.


\bibitem{2021_TSP_gxiqigao_RE_maximization}
L. You, J. Xiong, D. W. K. Ng, C. Yuen, W. Wang, and X. Gao, ``Energy efficiency and spectral efficiency tradeoff in RIS-aided multiuser MIMO uplink transmission,'' \emph{IEEE Trans. Signal Process.,} vol. 69, pp. 1407–1421, Dec. 2021.

%\bibitem{2021_TWC_Renzo_EE_channel_estimation}
%A. Zappone, M. Di Renzo, F. Shams, X. Qian, and M. Debbah, ``Overhead-aware design of reconfigurable intelligent surfaces in smart radio environments,'' \emph{IEEE Trans. Wireless Commun.,} vol. 20, no. 1, pp. 126-141, Jan. 2021.

\bibitem{2022_TGC_Ntontin_charging_power}
K. Ntontin et al., ``Wireless energy harvesting for autonomous reconfigurable intelligent surface,'' \emph{IEEE Trans. Green Commun. Netw.,} vol. 7, no. 1, pp. 114–129, Mar. 2023.


\bibitem{2021_OJCS_Lerosey_IRS_demo}
J.-B. Gros, V. Popov, M. A. Odit, V. Lenets, and G. Lerosey, ``A reconfigurable intelligent surface at mmwave based on a binary phase tunable metasurface,'' \emph{IEEE Open J. Commun. Society,} vol. 2, pp. 1055–1064, May 2021.

 %\bibitem{2021_TWC_Cui_power_consumption_channel_model}
%W. Tang et al., ``Wireless communications with reconfigurable intelligent surface: Path loss modeling and experimental measurement,'' \emph{IEEE Trans. Wireless Commun.,} vol. 20, no. 1, pp. 421-439, Jan. 2021.

\bibitem{2022_Arxiv_Cui_dynamic_power_consumption}
J. Wang et al., ``Reconfigurable intelligent surface: Power consumption modeling and practical measurement validation,'' \emph{IEEE Trans. Commun.,} early access, Mar. 28, 2024, doi: 10.1109/TCOMM.2024.3382332.



\bibitem{2023_Arxiv_Jin_dynamic_power_consumption}
J. Wang, W. Tang, S. Jin, X. Li, and M. Matthaiou, ``Static power consumption modeling and measurement of reconfigurable intelligent surfaces,'' in \emph{Proc. IEEE 31th Eur. Signal Process. Conf. (EUSIPCO),} Sept. 2023, pp. 890-894.

\bibitem{2023_Dai_PS-DPC}
Z. Li, J. Zhang, J. Zhu, S. Jin, and L. Dai, ``Enhancing energy efficiency for reconfigurable intelligent surfaces with practical power models,'' 2023, \emph{arXiv:2310.15901}. [Online]. Available: https://arxiv.org/abs/2310.15901.

\bibitem{2024_Jin_PS-DPC}
D. Xu, Y. Han, X. Li, J. Wang, and S. Jin, ``Energy efficiency optimization for a RIS-assisted multi-cell communication system based on a practical RIS power consumption model,'' \emph{Front. Inf. Technol. Electron. Eng.,} vol. 24, no. 12, pp. 1717-1727, Jan. 2024.

\bibitem{2022_TWC_IOS}
S. Zhang et al., ``Intelligent omni-surfaces: Ubiquitous wireless transmission by reflective-refractive metasurfaces,'' \emph{IEEE Trans. Wireless Commun.,} vol. 21, no. 1, pp. 219-233, Jan. 2022.

%\bibitem{2003_Rician_factor}
%Chia Leong Hong, I. J. Wassell, G. E. Athanasiadou, S. Greaves, and M. Sellars, ``Wideband tapped delay line channel model at 3.5GHz for broadband fixed wireless access system as a function of subscriber antenna height in suburban environment''. in \emph{Proc. 4th Int. Conf. Inf. Commun. Signal Process./4th Pac.-Rim Conf. Multimedia (ICICS-PCM)}, Dec. 2003, pp. 386-390.
%\bibitem{2020_ICC_Rayleigh channel}
%D. Kudathanthirige, D. Gunasinghe, and G. Amarasuriya, ``Performance analysis of intelligent reflective surfaces for wireless communication,'' in \emph{Proc. IEEE Int. Conf. Commun. (ICC)}, Jun. 2020, pp. 1-6.

\bibitem{2020_TWC_Zhang_PM_AO}
Q. Wu and R. Zhang, ``Beamforming optimization for wireless network aided by intelligent reflecting surface with discrete phase shifts,'' \emph{IEEE Trans. Wireless Commun.,} vol. 68, no. 3, pp. 1838-1851, Mar. 2020.

\bibitem{2016_JSAC_Yu_HBF}
X. Yu, J. -C. Shen, J. Zhang, and K. B. Letaief, ``Alternating minimization algorithms for hybrid precoding in millimeter wave MIMO systems,'' \emph{IEEE J. Sel. Topics Signal Process.,} vol. 10, no. 3, pp. 485-500, Apr. 2016.

\bibitem{2023_Arxiv_Schober_PM_GBD}
Y. Wu, D. Xu, D. W. K. Ng, R. Schober, and W. Gerstacker, ``Globally optimal resource allocation design for IRS-assisted multiuser networks with discrete phase shifts,'' in \emph{Proc. IEEE Int. Conf.
Commun. Workshops (ICC Workshops)}, May 2023, pp. 1216-1221.

\bibitem{2020_TWC_Larsson_sum rate maximization}
H. Guo, Y. -C. Liang, J. Chen and E. G, ``Weighted sum-rate maximization for reconfigurable intelligent surface aided wireless networks,'' \emph{IEEE Trans. Wireless Commun.,} vol. 19, no. 5, pp. 3064-3076, May 2020.

\bibitem{1972_JOTA_Geoffrion_GBD}
A. M. Geoffrion, ``Generalized Benders decomposition,'' \emph{J. Optim. Theory Appl.,} vol. 10, no. 4, pp. 237–260, Oct. 1972.

\bibitem{2023_TWC_Zhu_Han_GBD}
H. Huang, Y. Zhang, H. Zhang, Z. Zhao, C. Zhang, and Z. Han, ``Multi-IRS-aided millimeter-wave multi-user MISO systems for power minimization using generalized Benders decomposition,'' \emph{IEEE Trans. Wireless Commun.,} vol. 22, no. 11, pp. 7873-7886, Nov. 2023.


%\bibitem{1995_book_Floudas_v2_GBD}
%C. Floudas, \emph{Nonlinear and Mixed-Integer Programming—Fundamentals and Applications}. Oxford, U.K.: Oxford Univ. Press, 1995.

\bibitem{2020_CVX}
M. Grant and S. Boyd, ``CVX: Matlab software for disciplined convex programming, version 2.1,'' \emph{http://cvxr.com/cvx,} Jan. 2020.

\bibitem{Uniform Distribution of Sequences}
 L. Kuipers and H. Niederreiter, \emph{Uniform Distribution of Sequences.} New York: Wiley-Interscience, 1974.
 
\bibitem{STAR_IRS_phase_couple_2021}
J. Xu, Y. Liu, X. Mu, R. Schober, and H. V. Poor, ``STAR-RISs: A correlated T\&R phase-shift model and practical phase-shift configuration strategies,'' \emph{IEEE J. Sel. Topics Signal Process.,} vol. 16, no. 5, pp. 1097-1111, Aug. 2022.


\bibitem{2021_Zhu_WMMSE_HBF}
X. Zhao, T. Lin, Y. Zhu, and J. Zhang, ``Partially-connected hybrid beamforming for spectral efficiency maximization via a weighted MMSE equivalence,'' \emph{IEEE Trans. Wireless Commun.,} vol. 20, no. 12, pp. 8218-8232, Dec. 2021.

\bibitem{woodbury}
 H. Henderson and S. Searle, ``On deriving the inverse of a sum of matrices,'' \emph{Siam Rev.,} vol. 23, no. 1, pp. 53–60, Jan. 1981.

%\bibitem{2022_wU_BIOS_BF}
%Q. Wu, T. Lin, M. Liu, and Y. Zhu, ``BIOS: An omni RIS for independent reflection and refraction beamforming,'' \emph{IEEE Wireless Commun. Lett.,} vol. 11, no. 5, pp. 1062-1066, May 2022.

%\bibitem{2022_Lin_sparse_property_channel_estimation}
%T. Lin, X. Yu, Y. Zhu, and R. Schober, ``Channel estimation for IRS-assisted millimeter-wave MIMO systems: Sparsity-inspired approaches,'' \emph{IEEE Trans. Commun.,} vol. 70, no. 6, pp. 4078-4092, Jun. 2022.

%\bibitem{2016_Yu_CDA}
%F. Sohrabi and W. Yu, ``Hybrid digital and analog beamforming design for large-scale antenna arrays,'' \emph{IEEE J. Sel. Topics Signal Process.,} vol. 10, no. 3, pp. 501–513, Apr. 2016.

%\bibitem{2022_Mu_STAR-RIS}
%X. Mu, Y. Liu, L. Guo, J. Lin, and R. Schober, ``Simultaneously Transmitting and Reflecting (STAR) RIS Aided Wireless Communications,'' \emph{IEEE Trans. Wireless Commun.,} vol. 21, no. 5, pp. 3083-3098, May 2022.
\end{thebibliography}
\end{document}